\documentclass[useAMS,usenatbib]{mn2e}
\usepackage{epsf}
\usepackage{amsmath}
\usepackage{subfigure}

\newcommand{\kmsmpc}{\kms\;{\rm Mpc}^{-1}}

\newcommand{\hkpc}{h^{-1}{\rm kpc}}
\newcommand{\hmpc}{h^{-1}{\rm Mpc}}

\newcommand{\kms}{\;{\rm km}\,{\rm s}^{-1}}

\newcommand{\msun}{M_{\odot}}

\newcommand{\sfr}{\dot{M}_*}

\newcommand{\mlf}{\eta_{\rm{W}}}

\newcommand{\apjs}{ApJS}
\newcommand{\apj}{ApJ}
\newcommand{\apjl}{ApJL}
\newcommand{\aj}{AJ}
\newcommand{\araa}{ARA\&A}
\newcommand{\mnras}{MNRAS}
\newcommand{\prd}{PhRvD}

\newcommand{\hii}{\hbox{H\,{\sc II}}}

\newcommand{\fesc}{f_{\mathrm{esc}}}
\newcommand{\ncells}{N_{\mathrm{cells}}}

\newcommand{\nh}{n_{\mathrm{H}}}
\newcommand{\nhe}{n_{\mathrm{He}}}
\newcommand{\nhii}{n_{\mathrm{H\,II}}}
\newcommand{\nheii}{n_{\mathrm{He\,II}}}

\newcommand{\fJ}{f_{\mathcal{J}}}

\newcommand{\tes}{\tau_{\mathrm{es}}}
\newcommand{\zreion}{z_{\mathrm{reion}}}

\newcommand{\xmxv}{x_{\mathrm{M}}/x_{\mathrm{V}}}
\newcommand{\xhim}{x_{\mathrm{H\,I,M}}}
\newcommand{\xhiv}{x_{\mathrm{H\,I,V}}}
\newcommand{\xhiiv}{x_{\mathrm{H\,II,V}}}
\newcommand{\xhiim}{x_{\mathrm{H\,II,M}}}

\newcommand{\lmfp}{\lambda_{\mathrm{MFP}}}
\newcommand{\dxr}{\Delta x_{\mathrm{R}}}

\begin{document}

\title{ The Late Reionization of Filaments}

\author[astronomy]{Kristian Finlator, Feryal \"{O}zel, Romeel Dav\'e, \&
Benjamin D.\ Oppenheimer}

\maketitle

 \begin{abstract}
We study the topology of reionization using accurate three-dimensional
radiative transfer calculations post-processed on outputs from
cosmological hydrodynamic simulations.  In our simulations, reionization 
begins in overdense regions and then ``leaks" directly into voids, with
filaments reionizing last owing to their combination of high 
recombination rate and low emissivity.  This result depends on the 
uniquely-biased emissivity field predicted by our prescriptions for 
star formation and feedback, which have previously been shown to account 
for a wide array of measurements of the post-reionization Universe.  
It is qualitatively robust to our choice of simulation volume, ionizing 
escape fraction, and spatial resolution (in fact it grows 
stronger at higher spatial resolution) even though the exact overlap 
redshift is sensitive to each of these.  However, it weakens slightly 
as the escape fraction is increased owing to the reduced density contrast 
at higher redshift.  We also explore whether our results are sensitive 
to commonly-employed approximations such as using optically-thin 
Eddington tensors or substantially altering the speed of light.  Such 
approximations do not qualitatively change the topology of reionization.  
However, they can systematically shift the overlap redshift by up to 
$\Delta z\sim 0.5$, indicating that accurate radiative transfer is 
essential for computing reionization.  Our model cannot simultaneously 
reproduce the observed optical depth to Thomson scattering and 
ionization rate per hydrogen atom at $z=6$, which could owe to numerical 
effects and/or missing early sources of ionization.
\end{abstract}

 
\section{Introduction}
\label{sec:intro}
In the concordance $\Lambda$CDM cosmology, the sources that dominate 
cosmological reionization form predominantly in overdense regions.  
In the presence of an inhomogeneous intergalactic medium (IGM), the way 
in which ionization fronts (I-fronts) propagate into the IGM depends on 
the spatial distributions of the gas density and sources.  During the 
early stages of reionization, I-fronts proceed preferentially from the 
overdense knots, where sources form, toward underdense regions, where 
the hydrogen recombination rates are lower; this is referred to as 
``inside-out" (IO) reionization.  By contrast, in the final stages of 
reionization or within regions that have already reionized, reionization 
proceeds predominantly from voids towards overdensities; this is referred 
to as ``outside-in" (OI) reionization.  The way in which reionization 
proceeds from the initial IO phase to the final OI phase affects the
evolution of the size spectrum of ionized regions, leaving observable
imprints on the power spectrum of fluctuations in the 21cm 
background~\citep{fur05,mcq07} and the galaxy-21cm cross-correlation 
function~\citep{lid09}.  Additionally, the topology of reionization 
directly determines the dependence of the volume-averaged hydrogen 
recombination rate on the neutral hydrogen fraction through the
clumping factor because it determines the ``order" in which regions 
with different recombination rates are reionized.  Hence, it is an 
important ingredient in observational constraints on the strength of 
the ionizing background as well as in semi-analytic models of 
reionization.  

For these reasons, the topology of reionization has been the subject 
of numerous recent investigations.  In one of the pioneering radiative 
hydrodynamic simulations of cosmological reionization,~\citet{gne97} found 
that the clumping factor of ionized hydrogen increases monotonically with 
time.  While they did not specifically discuss the topology of reionization, 
this trend suggests a completely OI topology because the baryonic clumping 
factor is smaller in voids than in overdensities.  The approximations in this 
work effectively smoothed the ionizing background over a large cosmological
volume, and the resulting topology was probably an artifact of this 
treatment.  Similar results have been found in simulations that introduce
the ionizing background as a boundary condition~\citep{nak01}.  Indeed, 
soon afterwards,~\citet{gne00} found that a more accurate treatment for the 
spatial distribution of the ionizing sources yields a more IO-like 
reionization in the sense that the clumping factor of ionized hydrogen 
decreases monotonically in time.  

In the same year,~\citet{mir00} formulated an influential analytic
treatment for the final stages of reionization based on the idea
that the last regions to reionize are those that combine high density
with low emissivity.  This model also describes the evolution of
the ionization field within $\hii$~regions.  However, it does not
address the early stages of reionization, which are currently being
tackled by large-scale radiative transfer simulations that attempt
to model the expansion of I-fronts out of small haloes.  These 
calculations now consistently predict an IO topology in the sense 
that the ratio of the mass-averaged to the volume-averaged ionized 
hydrogen fraction $\xmxv$ is greater than unity at all
times~\citep{ili06a,ili07,tra07,lee08}.  Finally, a recent
semi-numerical work speculated that treating the spatial distribution
of the hydrogen recombination rate accurately could lead to a hybrid
scenario in which overdense regions ionize first, then underdense
regions, and then filaments~\citep{cho09}.  In summary, there is as 
yet no consensus on how the topology of reionization evolves from the 
Dark Ages until overlap.

We have recently developed an accurate technique for computing
cosmological reionization that uses a moment method to solve the
fully time-dependent radiative transfer equation~\citep{fin09}.
Here, we use this method to investigate the topology of reionization.
This work is complementary to previous studies in that we derive
the baryon density and emissivity fields from hydrodynamic simulations
whose baryonic physics treatments have been shown to reproduce a
wide range of observations of the post-reionization Universe,
including the high-redshift galaxy luminosity function~\citep{dav06}
and the metallicities of galaxies~\citep{fin08}, groups~\citep{dav08},
and the IGM~\citep{opp06,opp08,opp09}.  Of course, we are making
an assumption that the same emission sources and processes dominate
during reionization; in effect, we will test this assumption by
comparing to available observations.  Additionally, the present
work represents an improvement over the preliminary application
presented in~\citet{fin09} in three important respects: (1) we now
evolve the emissivity and density fields in time rather than assuming
them to be static; (2) the present simulations incorporate eight
times the cosmological volume while the underlying star formation
rates are computed using the same mass resolution; and (3) these
calculations begin at $z=14$ rather than $z=9$, so they account
more accurately for the impact of ionizing photons that were emitted
well before the hydrogen recombination time at the mean IGM density 
exceeded the Hubble time.

In Section~\ref{sec:sims}, we describe the cosmological simulation that 
we use as a basis for our post-processing radiative transfer calculations, 
and review our radiative transfer technique.  In Section~\ref{sec:topo}, 
we study the topology of reionization in our calculations.  In 
Section~\ref{sec:wmap}, we compare our results with constraints on the 
integrated optical depth to Thomson scattering and the volume-averaged 
ionizing emissivity at $z\sim6$.  We discuss our results in 
Sections~\ref{sec:disc}.  Finally, we summarize our findings in 
Section~\ref{sec:conc}.

\section{Simulations}
\label{sec:sims}

\subsection{Cosmological Simulation}
\label{sec:cosmo_sim}

We ran the cosmological hydrodynamic simulation that serves as an
input to our post-processing radiative transfer calculation using
our custom version of the parallel cosmological galaxy formation
code Gadget-2~\citep{spr02}.  This code uses an entropy-conservative
formulation of smoothed particle hydrodynamics (SPH) along with a
tree-particle-mesh algorithm for handling gravity.  It accounts for
photoionization heating starting at $z=9$ via a spatially uniform
photoionizing background~\citep{haa01}.  Gas particles undergo
radiative cooling under the assumption of ionization equilibrium,
where we account for metal-line cooling using the collisional
ionization equilibrium tables of \citet{sut93}.  Stars are formed
from dense gas via a subresolution multi-phase model that tracks
condensation and evaporation in the interstellar medium following
\citet{mck77}.  The model is tuned via a single parameter, the star
formation timescale, to reproduce the~\citet{ken98a} relation; see
\citet{spr03a} for details.  Self-enrichment of star-forming gas
and delayed feedback from old star particles are also treated.  We
account for galactic-scale superwind feedback using our momentum-driven
outflows with a normalization $\sigma_0=150\kms$.  For further
details on the physics treatments in the simulations,
see~\citet{opp06,opp08}.  Our simulation subtends a cubical volume
$16 \hmpc$ long on each side and uses $512^3$ dark matter and star
particles.  It assumes a cosmology where $\Omega_M=0.25$,
$\Omega_\Lambda=0.75$, $H_0=70\kmsmpc$, $\sigma_8=0.83$, and
$\Omega_b=0.044$.

\begin{figure*}
\centerline{
\setlength{\epsfxsize}{0.5\textwidth}
\centerline{\epsfbox{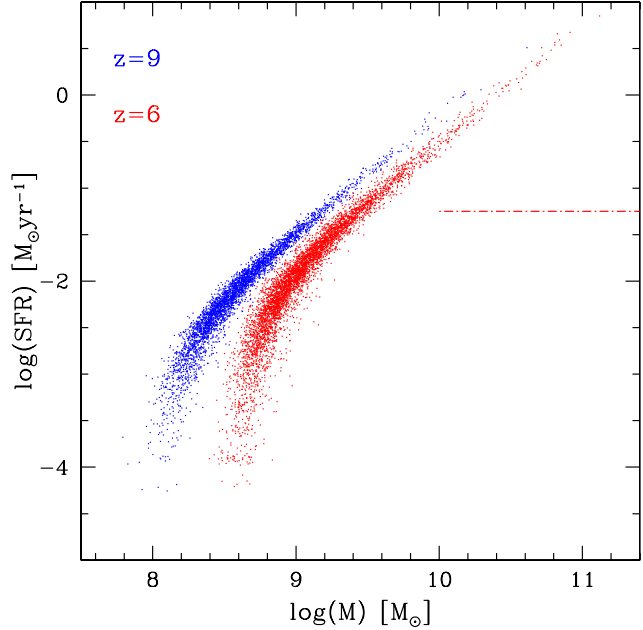}}
}
\caption{The simulated relationship between total halo mass and
instantaneous star formation rate at $z=9$ (upper, blue locus)
and $z=6$ (lower, red locus).  The simulated relationships roughly
follow SFR$\propto M^{1.3}$ at high masses, with normalizations and
low-mass cutoffs that evolve in time.  The red dot-dashed line 
indicates the current observational limit at $z=6$ (see text), which 
translates to a halo mass of $10^{9.5}\msun$.}
\label{fig:mhalo_sfr}
\end{figure*}

\begin{figure*}
\centerline{
\setlength{\epsfxsize}{0.5\textwidth}
\centerline{\epsfbox{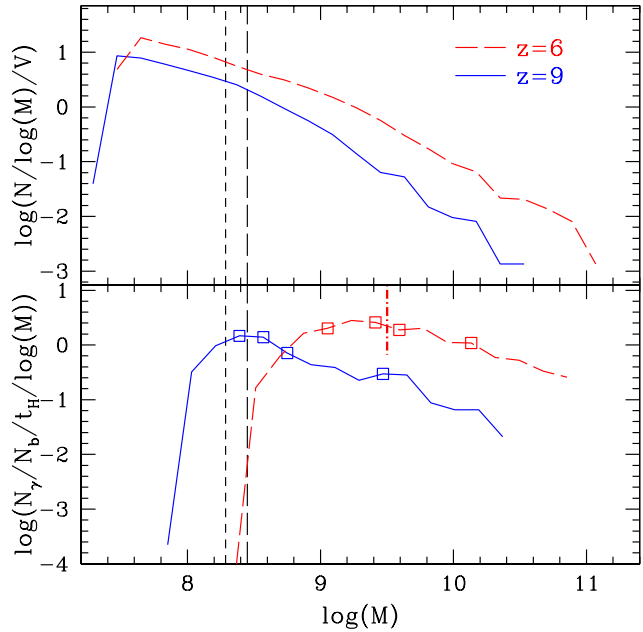}}
}
\caption{(Top) The simulated halo mass functions at $z=9$ (solid) and 
$z=6$ (dashed). (Bottom) The corresponding ionizing luminosity-weighted 
halo mass functions in units of ionizing photons per hydrogen atom per
Hubble time per mass bin.  The open squares indicate the lower mass 
limits above which (right to left) 0.2, 0.4, 0.6, and 0.8 of the total 
ionizing luminosity is emitted.  The vertical short dashed line 
denotes the 64 particle resolution limit for baryonic physics, indicating
that our simulation resolves star formation in haloes more massive than
$2\times10^8\msun$.  The vertical long-dashed line indicate the halo mass 
where gas infall is expected to be suppressed by 50\% at $z=6$~\citep{oka08}.
The vertical dot-dashed segment in the lower panel indicates the current 
observational limit at $z=6$ (see text).}
\label{fig:ion_mhalo}
\end{figure*}

We now briefly discuss how our mass resolution compares to the mass
scales that dominate reionization.  Haloes that are less massive
than the threshold for atomic cooling ($8.3\times10^7
(7/(1+z))^{1.5}\msun$) do not contribute significantly to reionization
owing to inefficient star formation~\citep{wis09} and the early
passage of Lyman-Werner photons~\citep{hai97,ahn09}, hence we are
interested only in more massive haloes.  Ideally, we would like to
resolve star formation in haloes down to roughly $2\times10^7\msun$,
the atomic cooling threshold at $z=14$.  In order to understand how
our ionizing emissivity field relates to the underlying halo
population, we have identified the dark matter haloes in our
cosmological volume at the representative redshifts $z=9$ and $z=6$
using a spherical overdensity algorithm that determines the smallest
radius out to which a halo's enclosed density falls below the virial
density.

We show in Figure~\ref{fig:mhalo_sfr} the predicted
relationship between halo mass $M$ and star formation rate $\sfr$.
For haloes above $10^9\msun$, both relationships follow a trend
$\sfr \propto M^{1.3}$.  The superlinear scaling results from our
momentum-driven outflows because the outflow mass loading factor
$\mlf$ scales inversely with the velocity dispersion, preferentially
suppressing the star formation rates of low-mass systems~\citep{opp06}.
At both redshifts, there is a turnover at low masses.  Defining the
turnover mass as the highest mass where the median $\sfr$ is below
50\% of the linearly extrapolated high-mass trend (extrapolated
from $M>10^9\msun$), we find that the turnover mass rises from
$10^{8.3}\msun$ at $z=9$ to $10^{8.7}\msun$ at $z=6$.  At $z\ga 9$,
the turnover owes to resolution effects, while at $z=6$ it owes
entirely to the suppression of cooling into low-mass
haloes~\citep[e.g.,][]{tho96,oka08} by the nascent ionizing
background~\citep{haa01}.  During the same interval, the normalization
of the $M-\sfr$ trend drops by roughly 0.25 dex owing to cosmological
expansion.

The red dot-dashed segment in Figure~\ref{fig:mhalo_sfr} indicates
the current observational limit at $z=6$ corresponding to 
$M_{\mathrm{UV,AB}}=-17.5$~\citep{bou07}.  We obtained this limit
using the relationship between star formation rate and rest-frame
1350 \AA~luminosity that arises in our hydrodynamic simulation at
$z=6$\footnote{Convolving our simulated stellar populations' star
formation histories and metallicities with the~\citet{bc03} models
and assuming no dust, this is given by $M_{\mathrm{UV,AB}} = -21.10
-2.88\log(\sfr)$, where $\sfr$ is in $\msun$ yr$^{-1}$.}.

The relationship in Figure~\ref{fig:mhalo_sfr} is a testable
prediction of our model because it is related to the rest-frame UV
luminosity function of Lyman-break galaxies at $z=6$.  It has
previously been shown (albeit with slightly different parameter
choices) that our model nicely reproduces the observed
normalization~\citep{dav06} and evolution~\citep{bou07} of the high
redshift galaxy luminosity function.  Hence the trends in
Figure~\ref{fig:mhalo_sfr} represent at least a plausible model for
the sources of ionizing photons at $z\ga 6$.

The top panel of Figure~\ref{fig:ion_mhalo} shows the simulated halo 
mass functions at the same redshifts.  The mass functions rise smoothly 
to lower masses until roughly $6.0\times10^7\msun$, which is the 20 dark 
matter particle resolution limit for haloes and lies close to the above 
target resolution.  Converting the dark matter haloes in the top panel 
into an ionizing emissivity field involves a host of additional 
assumptions regarding gas cooling and star formation.  The novelty of 
our method is that these processes are already treated in great detail 
by our hydrodynamic simulation, so that we are not obliged to make any 
additional assumptions when generating the emissivity field snapshots 
that we use for computing reionization save for the choice of ionizing 
escape fraction $\fesc$.  By convolving each halo's stellar population 
with the~\citet{bc03} population synthesis models assuming 
a~\citet{cha03} IMF, we have computed the corresponding ionizing 
luminosity-weighted halo mass functions.  We show these in the bottom 
panel in units of ionizing photons emitted per hydrogen atom per Hubble 
time per mass bin.  These curves show how haloes of different masses 
contribute to reionization, and the volume-averaged ionizing emissivity 
per hydrogen atom into the IGM is simply their integral multiplied by the 
escape fraction.   

At both redshifts, the ionizing emissivity per mass bin increases with
decreasing mass before turning over below $10^9\msun$.  The trend at high 
masses is flatter than the mass function (note that the y-axes in the 
bottom and top panels span the same dynamic range) because more massive 
haloes have higher star formation rates (Figure~\ref{fig:mhalo_sfr}), and 
it increases with decreasing mass because the high abundance of low-mass
haloes wins over their low individual star formation rates.

The low-mass turnover at each redshift reflects the behavior in 
Figure~\ref{fig:mhalo_sfr}.  At $z=9$, the turnover lies below 
$2\times10^8\msun$ and owes entirely to low resolution since the
threshold for resolved star formation rates is 64 star
particles~\citep{fin07}.  We crudely suggest this threshold with a
vertical short dashed line at 64 dark matter particle masses $=
10^{8.29}\msun$.  By $z=6$, the uniform ionizing background that
we employ in the hydrodynamic simulation~\citep{haa01} boosts the
minimum halo mass for efficient gas infall to
$\sim$~few$\times10^8\msun$~\citep{tho96,oka08}, which we indicate
with a vertical long dashed line.  This is well above the 64-particle
minimum for resolved star formation histories, hence the emissivity
field during the latter stages of reionization is very well-resolved.
Note that the suppression of star formation in low-mass haloes at 
late times bears some resemblance to self-regulated 
scenarios~\citep[e.g.,][]{ili07} although our prescription for 
regulating star formation is different.

The red vertical dot-dashed segment translates the observational limit
at $z=6$ from Figure~\ref{fig:mhalo_sfr} into halo mass, and the squares
indicate the minimum halo mass above which (from right to left) 20, 40, 
60, and 80\% of the ionizing photons are produced.  Comparing the dot-dashed
line and the boxes suggests that, if $\fesc$ is constant, then current 
observations are sensitive to only 50\% of the total ionizing luminosity 
density at $z=6$.

\subsection{Radiative Transfer Calculations}
\label{sec:radtx_sim}

We extract snapshots from this simulation at roughly 40-Myr intervals and 
map their baryonic density and emissivity fields onto a Cartesian grid 
for the radiative transfer integration.  We divide the total 
mass associated with SPH particles that lie near cell boundaries between 
the cells by summing incomplete gamma functions to their equivalent 
Plummer SPH smoothing kernels.  The gas temperature in each cell is the
mass-weighted average of the temperature of the gas particles that lie 
within the cell.  We omit from the grid those SPH particles whose density 
exceeds the threshold for star formation because their effect on the 
ionizing photons is implicitly accounted for in the choice of ionizing 
escape fraction.  We assume that all gas is completely neutral at $z=14$.  
We use radiative grids incorporating $32^3$, $48^3$ and $64^3$ cells, 
yielding radiative spatial resolutions $\dxr$ of 500, 333, and 250 
comoving $\hkpc$, respectively.  Poisson noise in the derived density 
fields is insignificant because, even at our highest resolution, there are 
on average $(512/64)^3 = 512$ gas particles within each cell.  

We compute each cell's emissivity by convolving its stellar populations with 
the~\citet{bc03} stellar population synthesis models, interpolating to the
correct age and metallicity for each star particle.  We tune the uniform 
ionizing escape fraction to 13\% so that the volume-averaged neutral fraction 
falls to roughly $10^{-3}$ at $z=6$.  We group all ionizing photons into a 
single frequency group at the hydrogen ionizing threshold.  

During the radiative transfer computation, we update the total baryon 
densities, emissivities, and temperatures using new snapshots from the
cosmological simulation while holding the radiative variables and ionization 
fractions constant.  The radiative transfer calculation accounts for 
cosmological expansion using the same cosmology as the hydrodynamical 
simulation.

Our radiative transfer calculations solve the moments of the radiative 
transfer equation in cosmological comoving coordinates.  After each 1-Myr 
timestep, we suspend the time-dependent integration of the moment equations 
and use a (time-independent) ray-tracing technique to compute the full 
angular dependence of the radiation field.  From this, we derive the updated 
Eddington tensor field, which is needed to close the moment hierarchy.  We 
use a number of optimizations to render this technique computationally 
feasible, as described in detail in~\citet{fin09}.  In brief, (i) we only 
update a computational cell's Eddington tensor when its photon number density 
has changed by more than $\fJ=5\%$ since its Eddington tensor was last updated; (ii) 
we terminate a ray-tracing computation when the optical depth between the source 
and the target cell along the ray exceeds 6; and (iii) we switch from using one 
to using two layers of periodic replica volumes in order to mimic the effects 
of periodic boundaries once the volume-averaged neutral fraction drops below 
50\%.  In~\citet{fin09}, we used an extensive suite of parameter convergence 
tests to verify that each of these optimizations introduces errors in the
ionization fractions of at most 10\%.

In this work, we introduce an additional optimization that enables
the code to transition smoothly between the different ray-tracing
approaches that are appropriate for the optically thick and thin
regimes.  Before cosmological $\hii$~regions overlap, the highly
inhomogeneous opacity and emissivity fields lead to strong spatial
variations in the Eddington tensor field.  These can only be treated
accurately by using ray-tracing to compute the optical depths from
every position to every source.  However, as reionization proceeds,
a growing fraction of the computational cells lie deep within large
$\hii$~regions where the Eddington tensors are dominated by other
sources located within the same $\hii$~region and towards which the
optical depth is negligible.  In these regions, simply assuming
that the optical depth to every source is zero becomes increasingly
valid, and the computationally-efficient optically thin approximation
is appropriate.  Optimization thus involves defining a criterion
to determine when a computational cell lies within a large
$\hii$~region.  We do this as follows: At all times, we store the
optically thin Eddington tensor field (that is, the result from
assuming that all optical depths are zero) in memory.  Wherever the
(time-dependent) photon number density exceeds the photon number
density from the (time-independent) optically thin approximation
by a factor of 2, we use the optically thin Eddington tensors rather
than performing ray-tracing.  We have used a parameter convergence
test similar to those in~\citet{fin09} to verify that this choice
incurs no more than 10\% errors in the ionization fractions while
speeding up the computation by roughly a factor of 2.

Through extensive testing, we have found that relaxing our accuracy 
criteria so that we recompute the Eddington tensors whenever the photon
number density changes by 10\% and switch to the optically thin 
approximation once the time-dependent photon density exceeds the 
optically thin value by 10\% introduces negligible changes into the 
topology and redshift of reionization (Figure~\ref{fig:zreion_sys}) 
although it does introduce typical errors of 20\% into the ionization 
states of individual cells.  We refer to this as the ``fast" scheme 
and employ it for tests of systematics.

We evolve the ionization field using either implicit or explicit
finite-differencing techniques depending on the local ionization 
timescale.  We do not evolve the gas temperature because we do not solve 
for the hydrodynamic response of the gas to the passage of ionization 
fronts.  At each timestep, we iterate between the updates to the 
ionization and radiation fields until they converge to $10^{-4}$. 

Our highest-resolution simulation required roughly 10,000 CPU hours on 
a shared-memory machine with 8 2.5-GHz Intel Xeon CPUs.  The overall
computation time for an individual reionization simulation scales with 
the total number of cells $\ncells$ approximately as $\ncells^{1.5}$, 
reflecting the fact that the time for the Eddington tensor updates 
scales in this way~\citep{fin09}.

\section{The Early Reionization of Voids}
\label{sec:topo}

\begin{figure*}
\centerline{
\setlength{\epsfxsize}{0.5\textwidth}
\centerline{\epsfbox{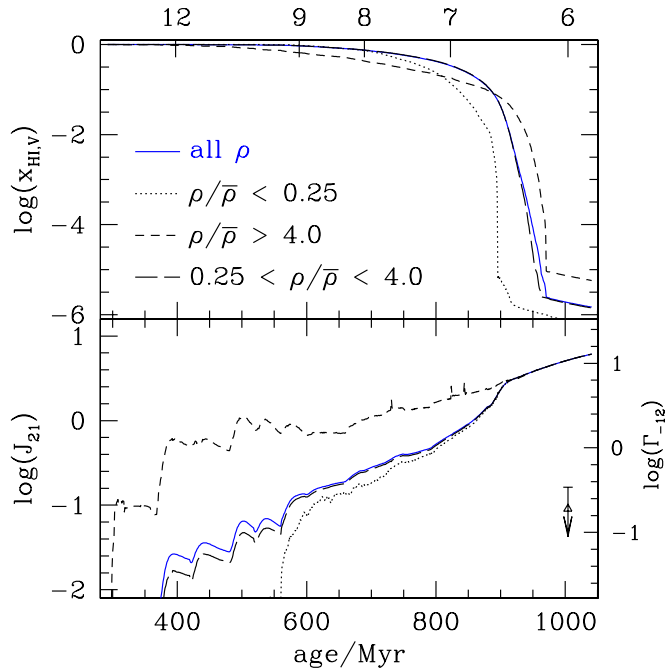}}
}
\caption{(Top) The volume-averaged neutral fraction in the IGM for
the $\dxr=250\hkpc$ calculation as a function of the age of the 
Universe (bottom axis) and redshift (top axis) overall and for three 
bins in overdensity.  (Bottom) The mean intensity of the ionizing 
background at 912 \AA~for the same density bins.  The ``notches" at 
roughly 40-Myr intervals (especially clear in the bottom panel) 
indicate the timesteps where the ionization and density fields are 
updated during the integration.  The triangle indicates the 
observed upper limit~\citep{bol07}.  Voids reionize before 
mean-density regions owing to the spatial distribution of 
recombination rates and ionizing intensities.}
\label{fig:xHI_vs_t}
\end{figure*}

\subsection{The Volume-Averaged Neutral Fraction} \label{sec:xHI}

In the top panel of Figure~\ref{fig:xHI_vs_t}, we show how the 
volume-averaged neutral fraction evolves in time at our highest
spatial resolution ($\dxr=250\hkpc$).  The solid curve gives the 
average over our entire simulation volume while the dotted, 
short-dashed, and long-dashed curves correspond to underdense, 
overdense, and mean-density regions, respectively.  Note that 
using different ranges does not change the qualitative results.  
Hereafter, we refer to underdense regions as ``voids" and 
mean-density regions as ``filaments". 

The first thing to notice is that overlap (defined as the point
where the volume-averaged neutral fraction $\xhiv$ drops below
$10^{-3}$) occurs by $z=6$.  This is a nontrivial 
result, given that our simulations have previously been shown to
reproduce a wide array of observations of the post-reionization
Universe (Section~\ref{sec:radtx_sim}) and that we have not
introduced any new baryonic physics in order to bring about 
reionization other than the (reasonable) choice of ionizing 
escape fraction.  This supports previous suggestions that 
reionization could have been dominated by ordinary (that is, 
not Population III) star formation~\citep[e.g.,][]{shu08}.

Examining the reionization history in different density bins, we
find that overdense regions reionize first because they host the 
bulk of the ionizing sources.  Around $z=9$, photons bypass the 
filaments and flow into the voids, which ionize rapidly owing to 
their low recombination rates.  Filaments ionize 
last because they possess fewer sources than overdense regions 
but higher recombination rates than voids.  We refer to this 
topology, in which the voids are ionized significantly before the 
filaments, as the ``inside-outside-middle" topology (IOM).
Following overlap, the neutral hydrogen fraction increases 
with overdensity as expected in the presence of a uniform 
ionizing background.

In the bottom panel of Figure~\ref{fig:xHI_vs_t}, we show how the 
volume-averaged mean ionizing intensity evolves in the same regions.  
Overdense regions see the strongest ionizing background owing to their 
proximity to sources.  The mean intensity in filaments always
represents an average of the intensities in reionized regions close to 
sources and self-shielded regions farther away (Figure~\ref{fig:maps}), 
but broadly it lies between the intensities in overdense and underdense 
regions.  The intensity in voids is negligible until the first I-fronts
bypass filaments at $z=9$.  Afterwards, voids ionize rapidly owing to 
their low recombination rates and their mean intensity rapidly reaches
the volume average.  Following overlap at $z\approx 6$, the ionizing 
background is very nearly homogeneous.

\begin{figure*}
\centerline{
\setlength{\epsfxsize}{0.5\textwidth}
\centerline{\epsfbox{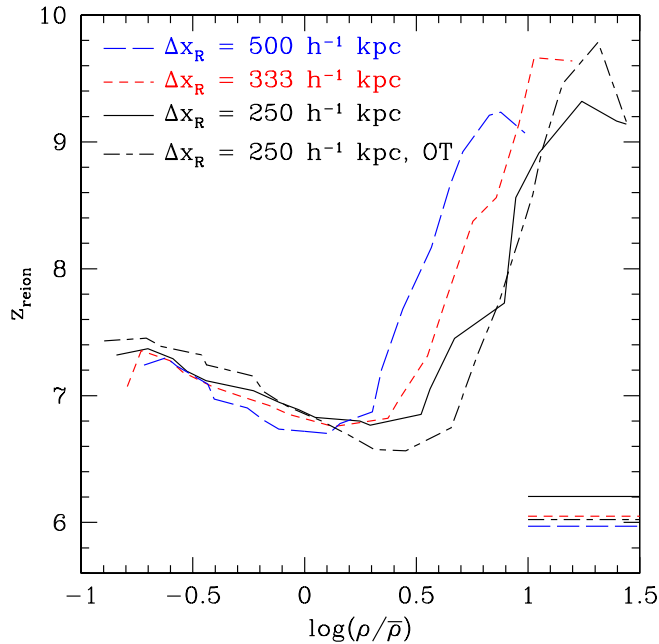}}
}
\caption{The median redshift of reionization as a function of overdensity
$\rho/\bar{\rho}$ (curves) and overlap redshift (horizontal lines) for
different choices of $\dxr$.  The short dashed - long dashed curve results 
from using $\dxr=250\hkpc$ along with optically thin (OT) Eddington 
tensors.  Broadly, the early reionization of voids is robust to the choice 
of $\dxr$ as well as the accuracy of the Eddington tensors.  Increasing 
the spatial resolution allows us to resolve both the high recombination 
rates in overdense regions and the tendency for photons to bypass 
filaments and reionize voids early.  The OT approximation delays overlap
by $\Delta z=0.2$ by delaying reionization in filaments.
}
\label{fig:zreion}
\end{figure*}

\begin{figure*}
  \subfigure[Overdensity $\rho/\bar{\rho}$ at $z=6.49$.]{
    \label{subfig:rho}
    \setlength{\epsfxsize}{0.48\textwidth}
    \epsfbox{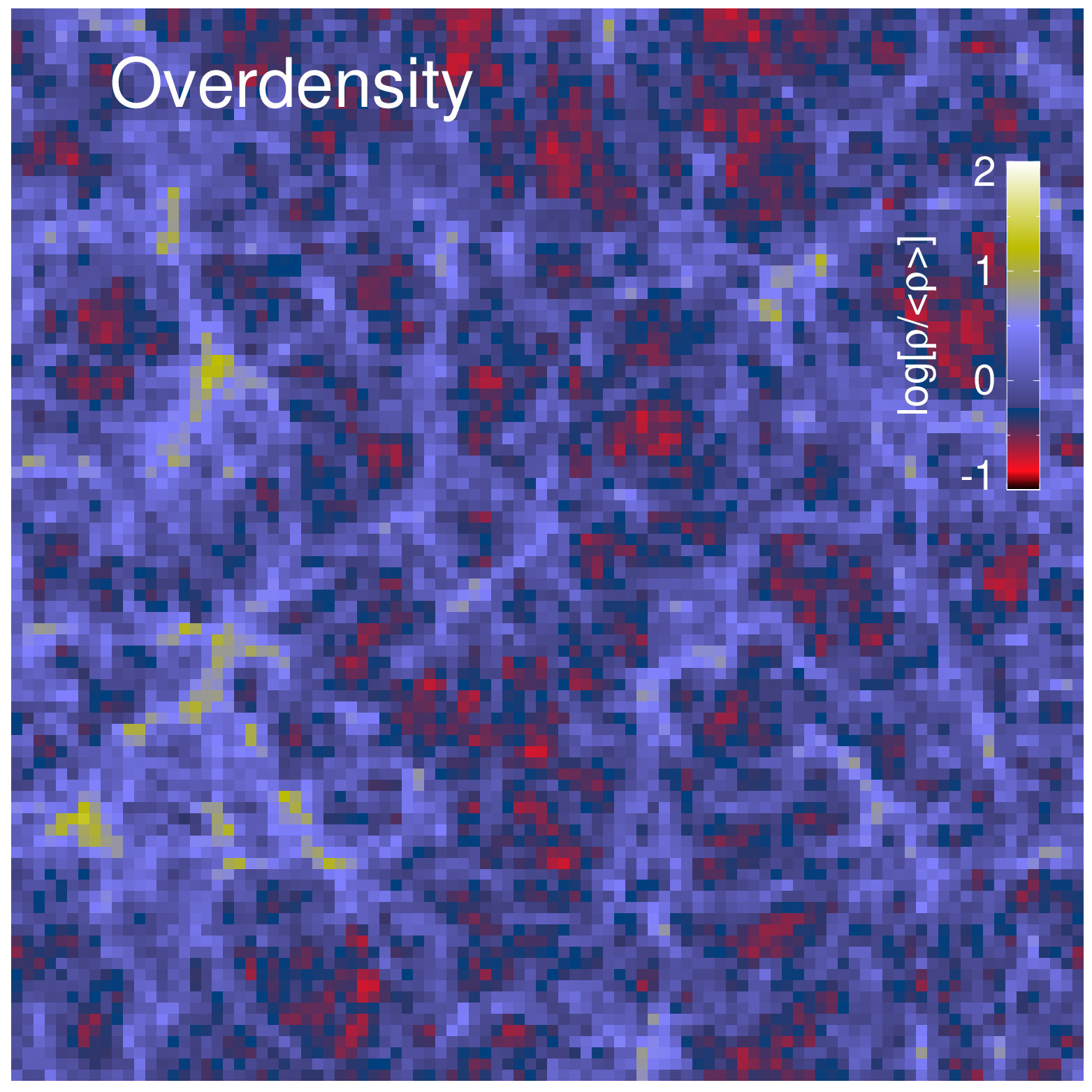}}
  \subfigure[Redshift of reionization, defined as the redshift at which the
    hydrogen neutral fraction first dips below $10^{-3}$.]{
    \label{subfig:zreion}
    \setlength{\epsfxsize}{0.48\textwidth}
    \epsfbox{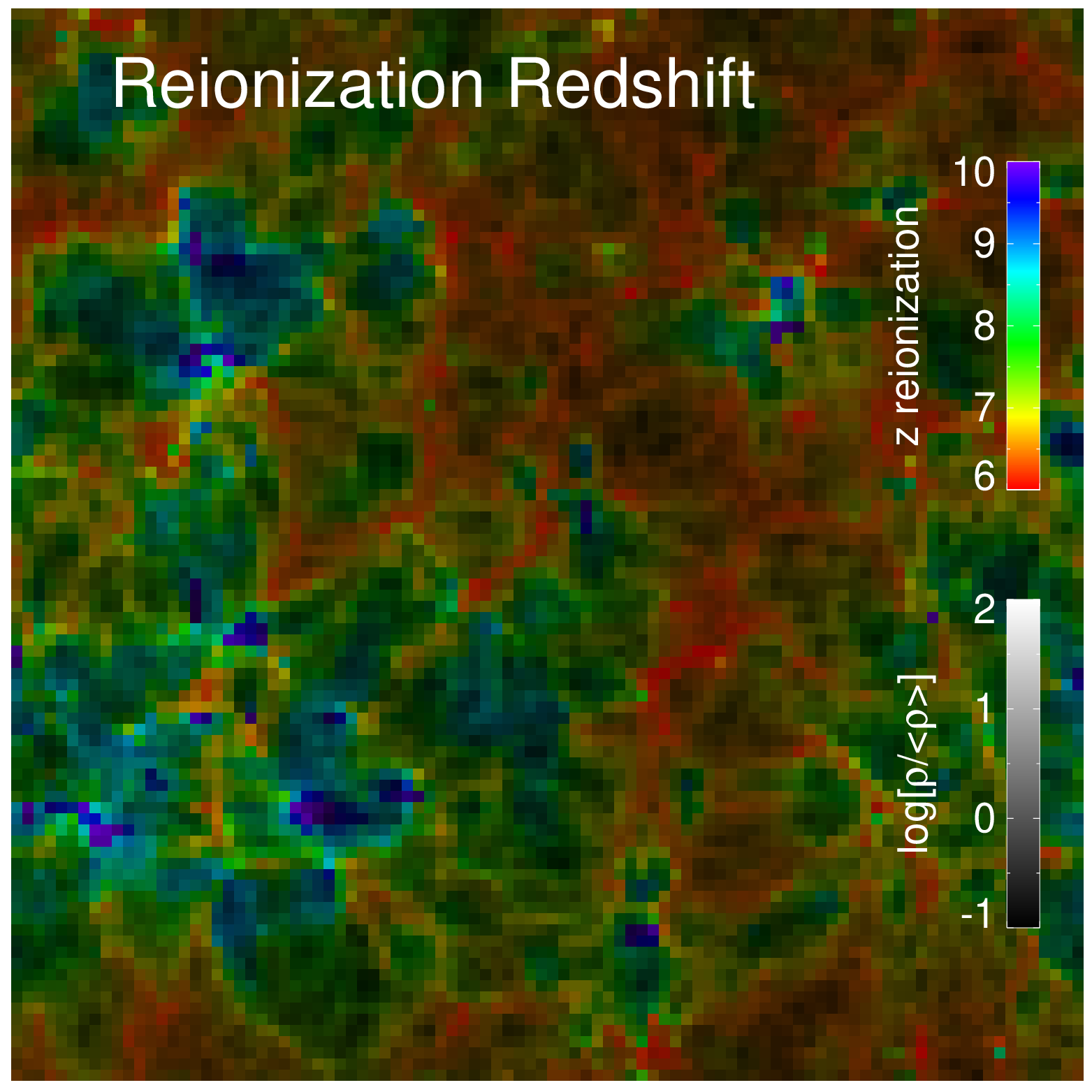}}\\ \vspace{-0.25in}
  \subfigure[Emissivity, $z=7.21$ ($\xhiv=0.5$).]{
    \label{subfig:eta_z721}
    \setlength{\epsfxsize}{0.31\textwidth}
    \epsfbox{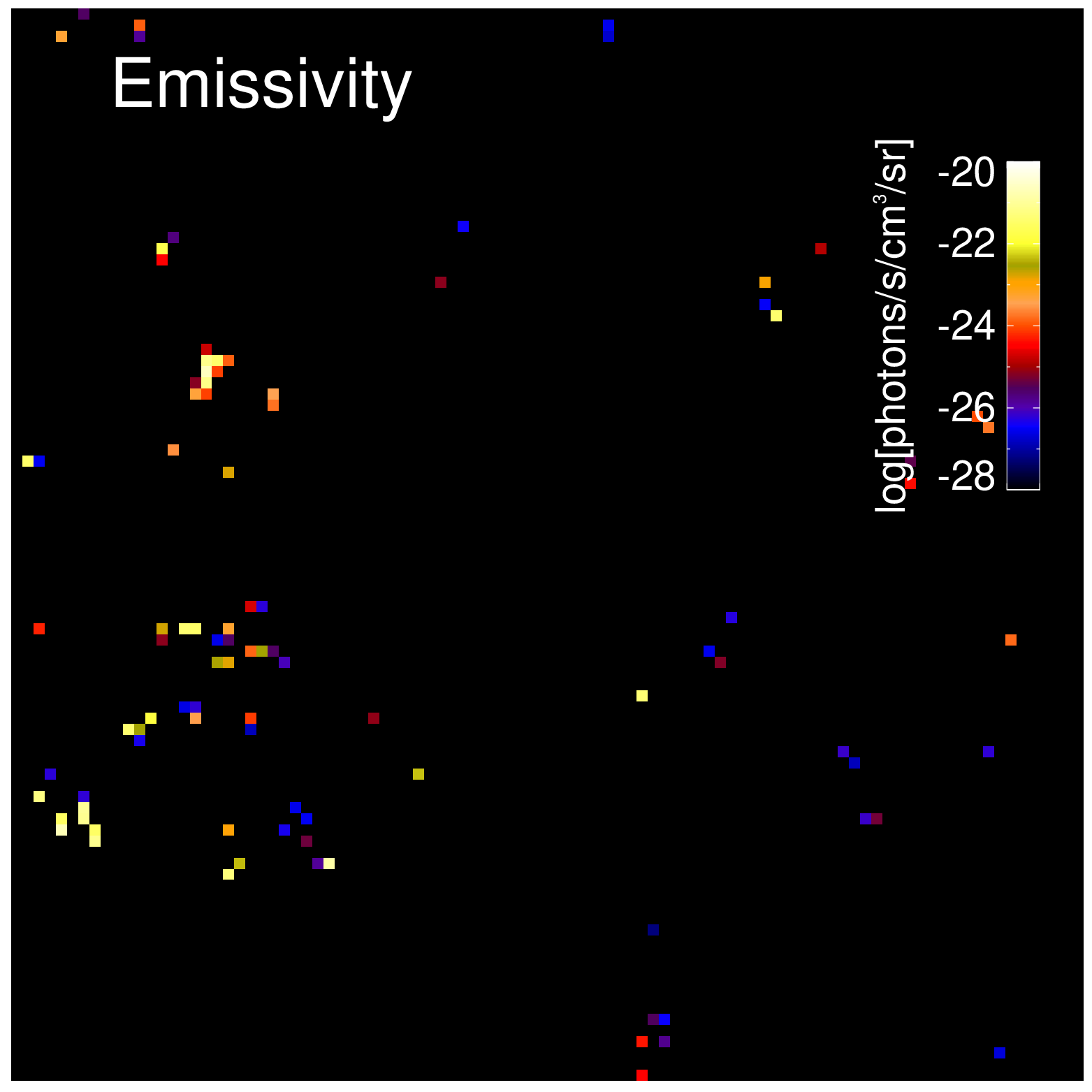}}
  \subfigure[$\xhim$ at $z=7.21$.]{
    \label{subfig:xhi_z721}
    \setlength{\epsfxsize}{0.31\textwidth}
    \epsfbox{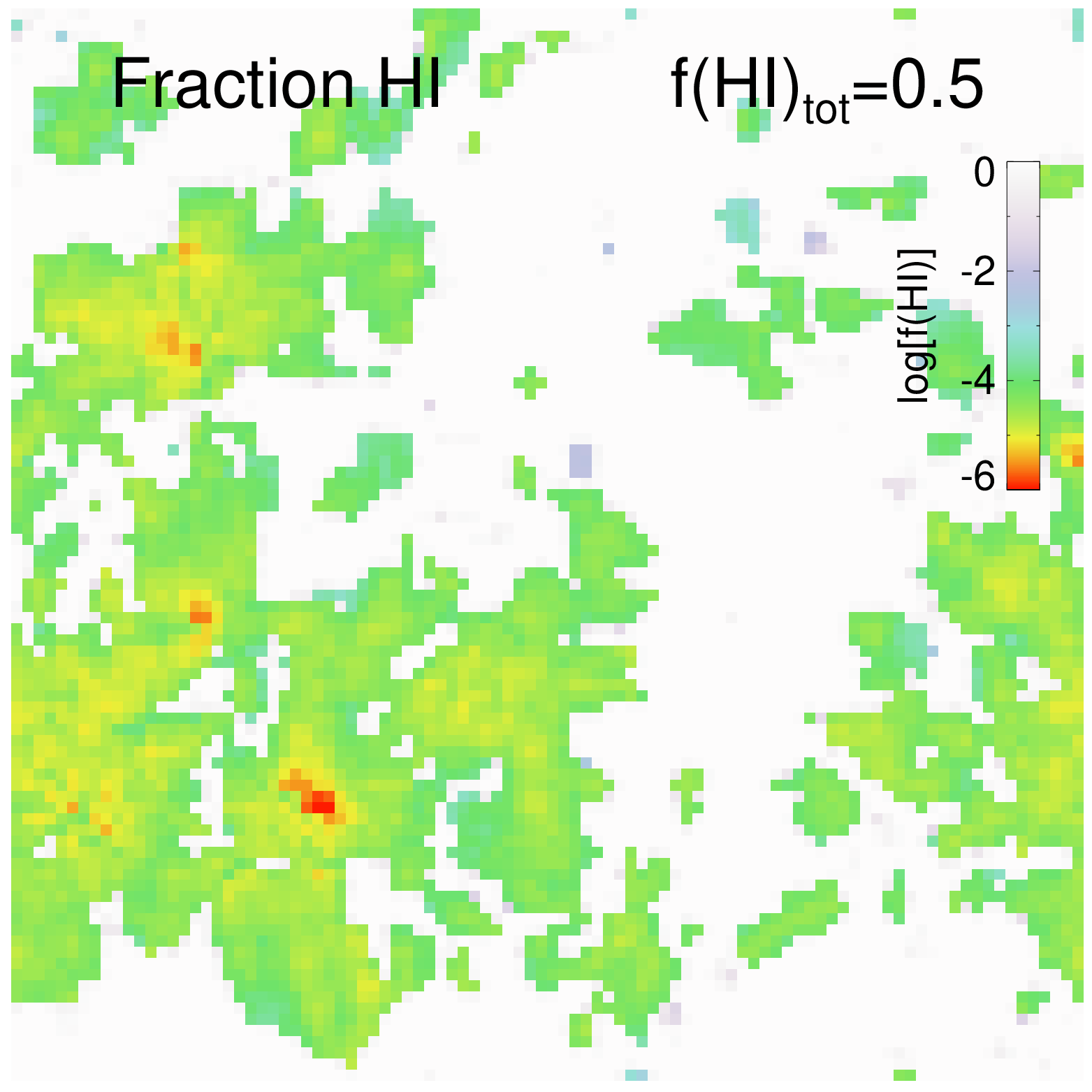}}
  \subfigure[Photon density at $z=7.21$.]{
    \label{subfig:J_z721}
    \setlength{\epsfxsize}{0.31\textwidth}
    \epsfbox{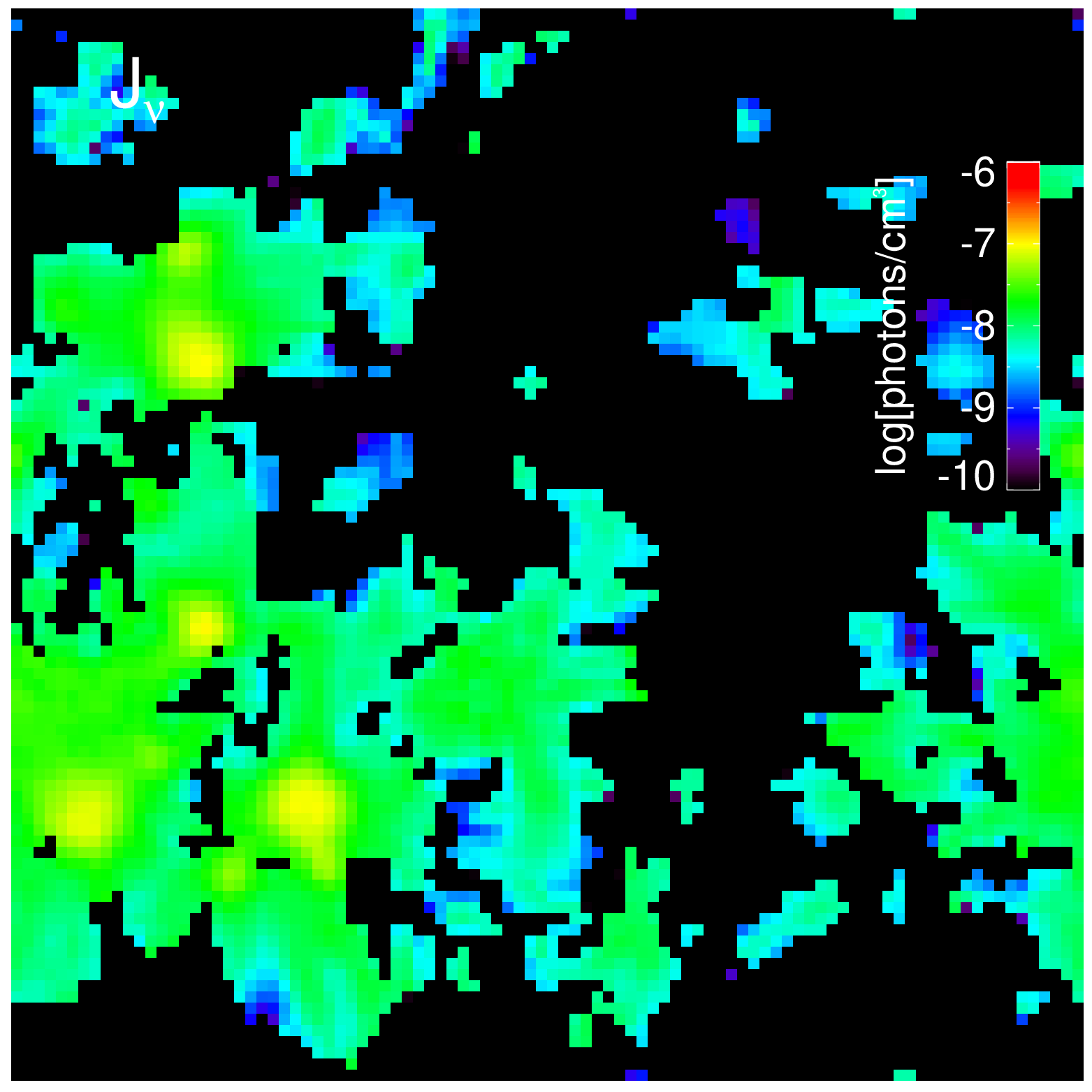}}\\ \vspace{-0.25in}
  \subfigure[Emissivity, $z=6.49$ ($\xhiv=0.1$).]{
    \label{subfig:eta_z649}
    \setlength{\epsfxsize}{0.31\textwidth}
    \epsfbox{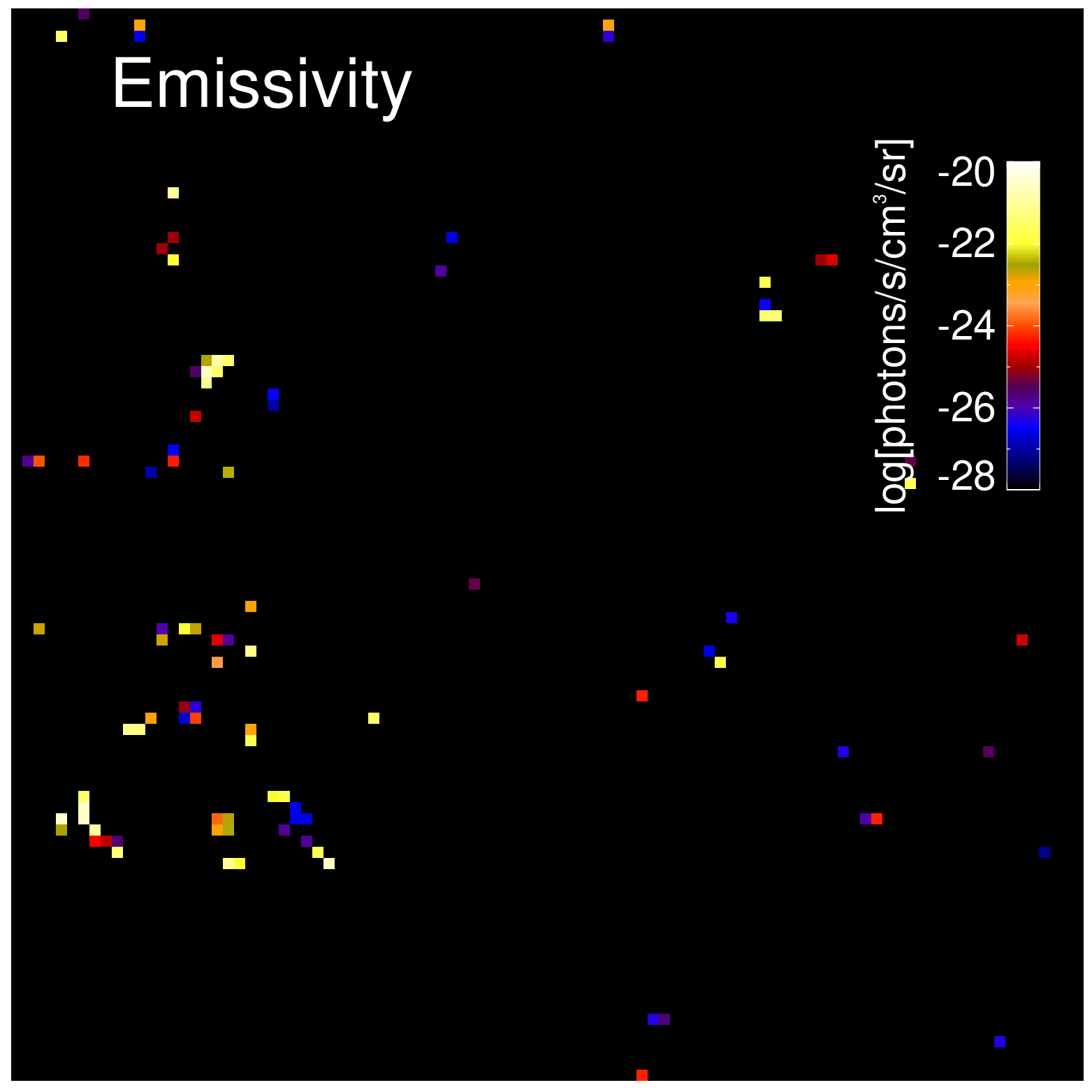}}
  \subfigure[$\xhim$ at $z=6.49$.]{
    \label{subfig:xhi_z649}
    \setlength{\epsfxsize}{0.31\textwidth}
    \epsfbox{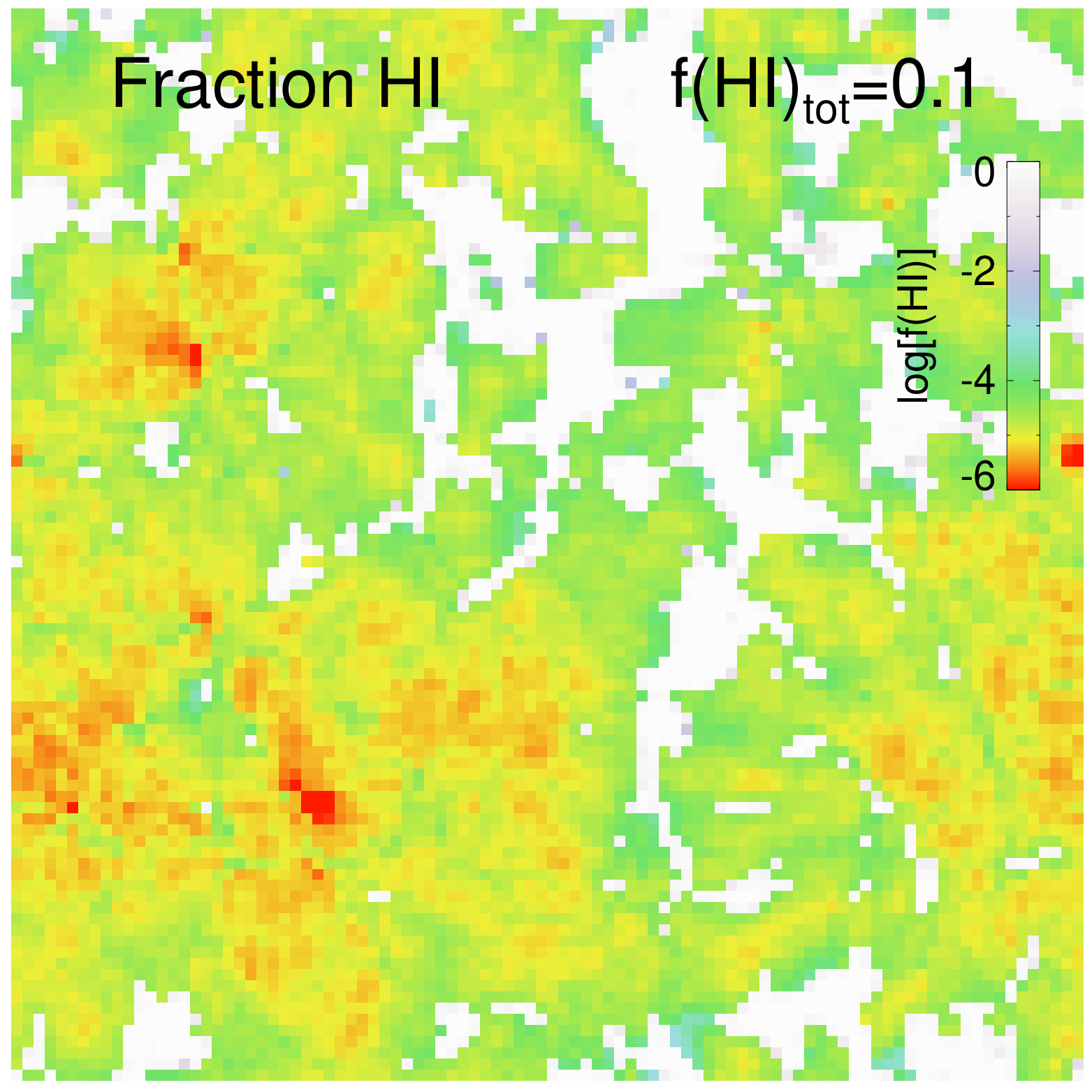}}
  \subfigure[Photon density at $z=6.49$.]{
    \label{subfig:J_z649}
    \setlength{\epsfxsize}{0.31\textwidth}
    \epsfbox{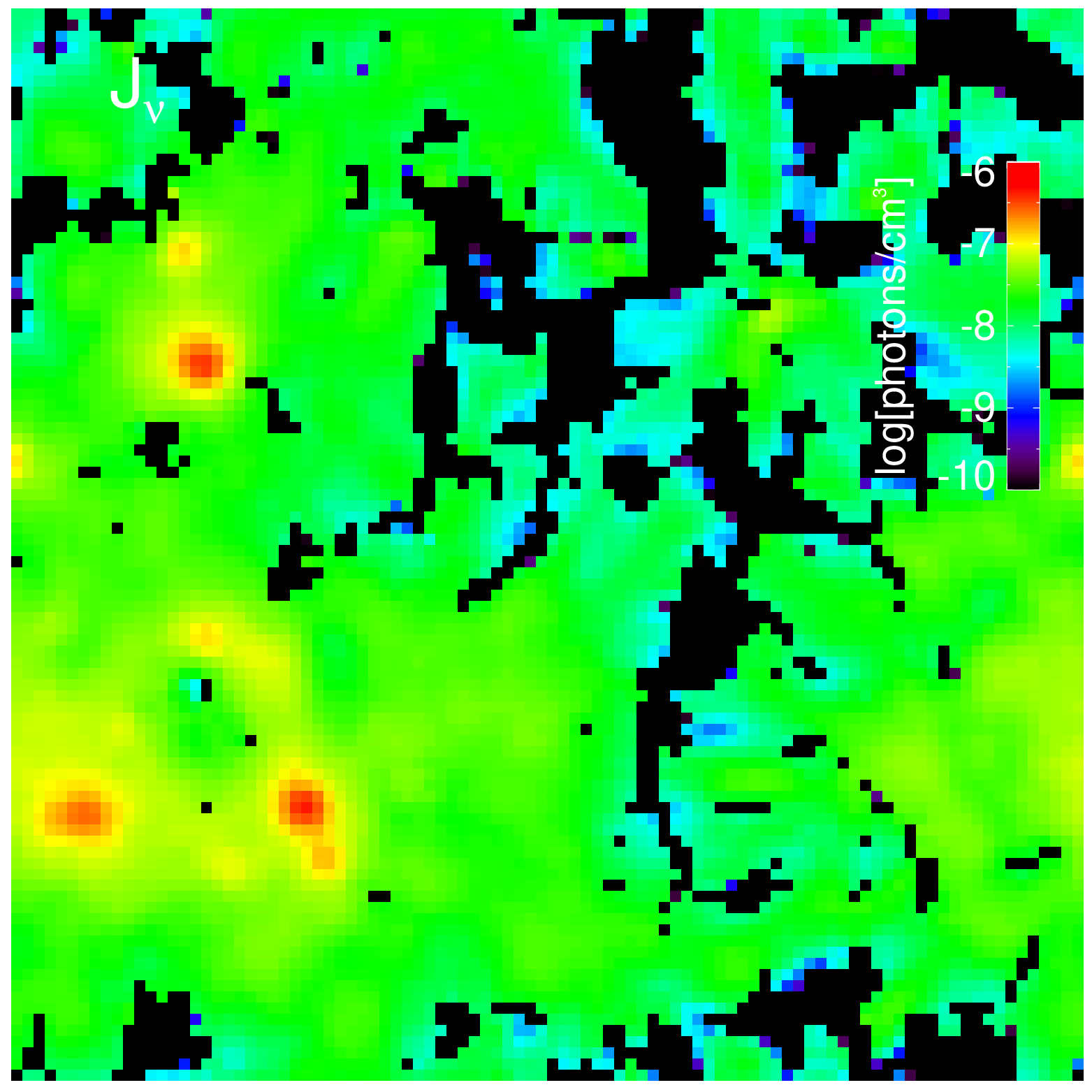}} \vspace{-0.15in}
  \caption{Maps of overdensity (a), redshift of reionization (b), ionizing 
    photon emissivity (c,f), neutral hydrogen fraction (d,g), and the 
    intensity of the ionizing background (e,h) in a volume slice $16\hmpc$ to 
    a side and $250\hkpc$ thick.  Figures c--h are taken at two representative 
    redshifts as indicated.  The knots between filaments tend to host sources, 
    but the filaments themselves have negligible emissivity, with the result 
    that they reionize last (red color in panel b).}
  \label{fig:maps}
\end{figure*}

In Figure~\ref{fig:zreion}, we show how the median redshift of 
reionization varies with overdensity and $\dxr$.  We constructed this 
figure by determining the first redshift at which each cell's neutral 
hydrogen fraction drops below $10^{-3}$.  We define each cell's baryonic 
overdensity at the final timestep ($z=5.71$).  Broadly, our finding that 
reionization occurs sooner in voids than in filaments is insensitive to 
$\dxr$.  Decreasing $\dxr$ delays reionization in overdense regions by
enhancing I-front trapping while accelerating the reionization of voids.  
Defining the overlap epoch as the redshift at which the volume-averaged
neutral fraction drops below $10^{-3}$, we find that increasing the
spatial resolution from $\dxr=500$ to $250\hkpc$ causes overlap to occur
earlier by $\Delta z=0.2$.  We will argue in Section~\ref{sec:mfp} that 
both of these effects owe to more effective ``tunneling" of photons 
through soft spots in the IGM at higher spatial resolution.  

The solid and short dashed - long dashed curves in Figure~\ref{fig:zreion}
show how reionization proceeds when we use Eddington tensors that are 
derived accurately versus in the optically thin limit, respectively.  
Using optically thin Eddington tensors systematically accelerates 
reionization in voids while delaying it in filaments. Overdense regions 
remain unaffected.  The artificially late reionization in filaments 
delays the overlap epoch by $\Delta z = 0.2$.  Note that this 
constitutes the first evaluation of the consequences of incorporating 
inaccurate Eddington tensors into calculations of cosmological 
reionization.

In order to show more intuitively how the topology in 
Figure~\ref{fig:zreion} arises, we show in Figure~\ref{fig:maps} maps 
of overdensity (a), ionizing emissivity (c,f), neutral hydrogen 
fraction (d,g), and the intensity of the ionizing background (e,h) at 
two representative redshifts as well as the redshift of reionization 
(b) in the same thin slice through our cosmological volume (see 
caption for details).  We note that the simulation from which we 
extract these maps is identical to the runs in Figure~\ref{fig:zreion} 
except that it uses $96^3$ radiative transfer cells for a radiative 
transfer resolution of $\dxr=177 \hkpc$.  Because this run became 
prohibitively slow after the neutral fraction dropped below 
$\xhiv < 10$\%, we completed its final stages using optically thin 
Eddington tensors.  This impacts the topology of reionization 
negligibly---in fact, we found that this simulation continues all of
our trends with respect to resolution---but for consistency we omit 
it from the other figures in this paper.

Panel~\ref{subfig:zreion} can be compared to Figure 1 of~\citet{tra08},
who found a purely IO reionization topology.
Comparing panels a, c, and f reveals that, while sources lie at the 
intersections between filaments as expected, they do not trace the 
filaments smoothly out to the mean density owing to the low-mass cutoff 
in the halo mass to light ratio (Figure~\ref{fig:ion_mhalo}).  This 
suggests that, for the most part, the filaments cannot self-ionize.  
Panels d, e, g, and h show that some filaments do ionize rapidly owing 
to proximity to sources, but at distances of more than $\sim$a few 
$\hmpc$ from sources, filaments remain self-shielded even after most 
of the voids have already reionized.  The tendency for most filaments 
to reionize late gives rise to our characteristic IOM reionization topology.

A close examination of the emissivity maps (panels~\ref{subfig:eta_z721}
and~\ref{subfig:xhi_z649}) shows that our simulated emissivity
varies rapidly with position.  This is a consequence of the fact
that the ionizing luminosity of a single stellar population varies
by 5 orders of magnitude during its first 100 Myr.  Our limited
mass resolution leads to shot noise in the number of star particles
per grid cell whose age lies within this range, which in turn gives
rise to the noisy emissivity field.  To gauge the impact of this,
we tried computing and using instantaneous star formation rates
directly from gas particles as the source of ionizing flux, rather
than the star formation history from star particles.  This would be a 
more appropriate approach in the limit that all ionizing photons are 
emitted instantaneously.  We left the escape fraction constant 
and tuned the ionizing luminosity  per unit star formation rate to 
$2.45\times10^{53}$ s$^{-1}$ per $\msun$ yr$^{-1}$ in order to match 
the volume-averaged emissivity resulting from the star 
particles\footnote{This is 2.64 times the~\citet{ken98b} relation.  
The extra ionizing luminosity can be attributed to the low metallicity 
at $z\geq6$.}.  We show the resulting reionization topology and overlap 
redshift with the dotted green lines in Figure~\ref{fig:zreion_sys}.  
Comparing these trends with the solid black lines, which reflect our 
fiducial emissivity prescription, we find that the two approaches yield 
nearly identical results.  This indicates that our overall results are not 
sensitive to stochasticity in the star formation algorithm.

\begin{figure*}
\centerline{
\setlength{\epsfxsize}{0.5\textwidth}
\centerline{\epsfbox{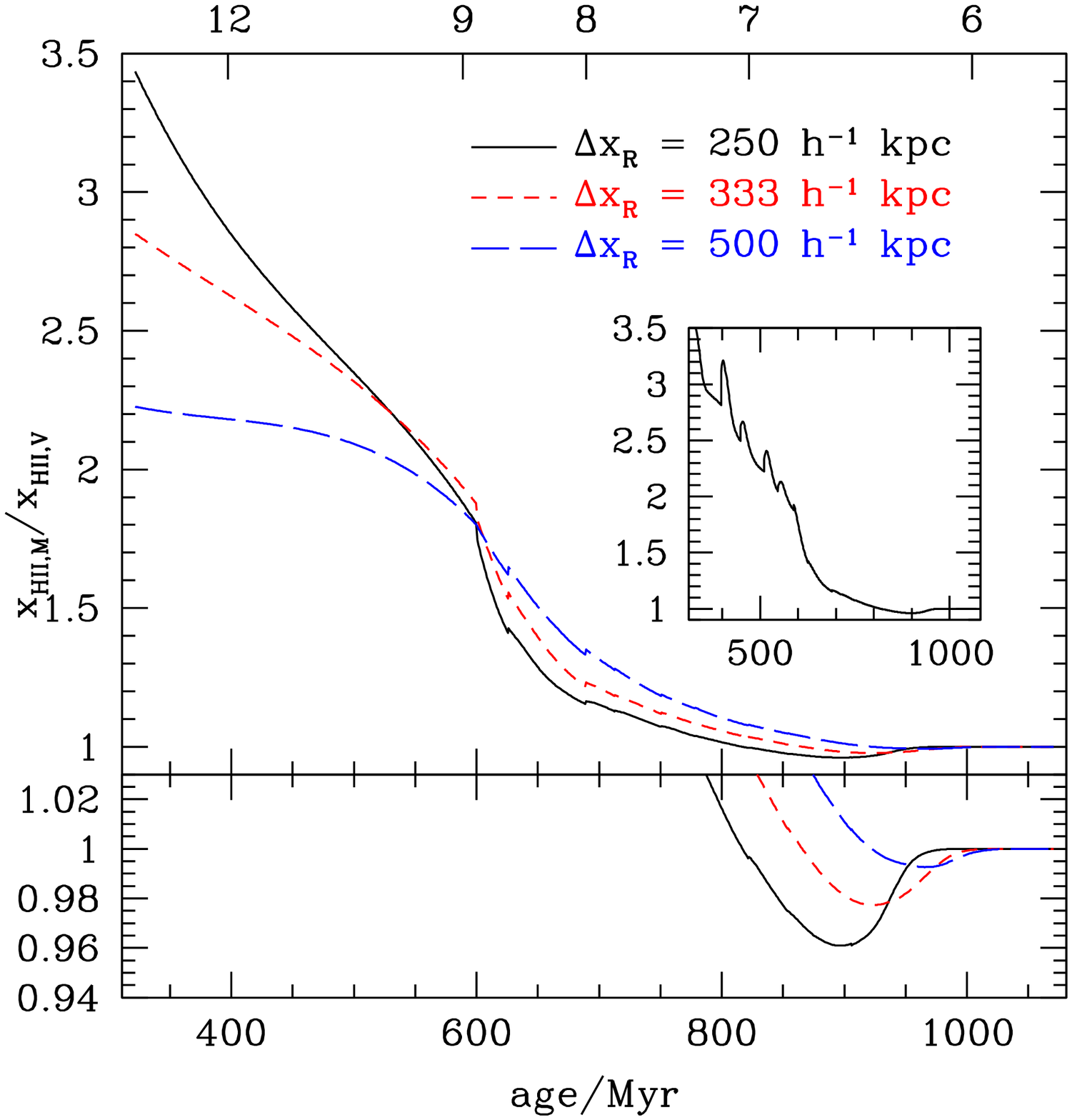}}
}
\caption{The ratio of the mass-weighted to volume-weighted ionized 
hydrogen fractions $\xmxv$ as a function of the age of the Universe 
(lower axis) and redshift (top axis).  The top panels shows how 
$\xmxv$ drops more rapidly at higher spatial resolution owing to 
more efficient breakout of I-fronts into voids.  We use polynomial
fits to the simulated trends for $z\geq9$.  The inset panel shows the
full simulated result for $\dxr=250\hkpc$, where the jumps owe to the
low time resolution in our set of precomputed density and emissivity 
fields.  The bottom panel expands the y-axis from the top panel about 
$\xmxv=1$ in order to show how $\xmxv$ drops below unity well before 
reionization completes, with the effect growing stronger at higher 
resolution.}
\label{fig:xmxv}
\end{figure*}

\subsection{The Ratio of Ionized Fractions} \label{sec:xmxv}

Another way to study the topology of reionization is to consider the 
ratio of the mass-averaged to the volume-averaged ionized hydrogen 
fractions, $\xmxv$, which can be thought of as the average density of 
ionized regions in units of the mean density~\citep{ili06a}.  For 
reionization in a homogeneous IGM, $\xmxv=1$ at all times.  In an 
inhomogeneous IGM, OI reionization implies that $\xmxv \leq 1$ because 
ionizing voids increases $\xhiiv$ more rapidly than $\xhiim$.  Similarly, 
IO reionization implies that $\xmxv \geq 1$ because ionizing 
overdense regions increases $\xhiim$ more rapidly than $\xhiiv$.  In 
Figure~\ref{fig:xmxv}, we show how $\xmxv$ evolves in time for three 
different choices of $\dxr$ in the radiation grid.  For the period
$z\geq9$, we smooth the simulated results with cubic polynomial fits 
in order to suppress artifacts from low time resolution in our set of 
precomputed density and emissivity grids.  The inset panel shows the 
original trend for the $\dxr=250\hkpc$ calculation.  

Broadly, $\xmxv$ evolves from $>1$ at early times when the ionized 
volume fraction remains low, to $<1$ when the voids reionize, to unity 
once the majority of the universe has reionized~\citep[see also][]{ili07}.  
The early evolution is as expected if reionization begins earlier in 
overdense regions than in the Universe as a whole~\citep{gne00,ili06a}, 
but the tendency for $\xmxv$ to drop below unity significantly before 
reionization completes has not been discussed previously.  The evolution 
of $\xmxv$ is sensitive to the choice of $\dxr$: Increasing spatial 
resolution increases $\xmxv$ at earlier times while decreasing it at late 
times.  Physically, higher spatial resolution allows the simulation to 
better resolve the high recombination rates in the overdense regions 
where reionization begins, leading to more effective trapping of I-fronts 
at early times.  At late times, increasing the resolution allows the 
simulation to resolve the tendency for I-fronts to ``leak out" through 
the porous IGM into voids.  As the bottom panel of Figure~\ref{fig:xmxv} 
indicates, this leaking effect is so efficient that, at our highest
spatial resolution, $\xmxv$ drops below unity around the time when 
$\xhiv$ drops below 50\% (at $z\approx7$).  In other words, {\it by the 
time reionization is halfway complete, the average ionized region is 
underdense}.  This is a direct consequence of the late reionization of 
filaments.

\subsection{The Mean Free Path of Ionizing Photons} \label{sec:mfp}

The tendency of ionizing photons to stream preferentially in directions
where the IGM is less dense naturally leads to the formation of ionized 
``tunnels" that connect sources with voids.  If this tunnel formation 
indeed dominates the topology of reionization, and if the tunnels are 
small compared to $\dxr$, then the mean free path of ionizing photons 
$\lmfp$ from sources should increase with decreasing $\dxr$.  This is 
because, as the tunnelling process is better-resolved, an increasing 
fraction of ionizing photons travel directly from sources through tunnels 
into voids rather than being artificially absorbed nearby.  To test this 
idea, we have computed $\lmfp$ as a function of redshift by casting rays 
from each source (where a source is a computational cell with nonzero 
emissivity) in 1280 directions that uniformly sample the unit sphere and 
then computing the distance travelled until the optical depth exceeds 6.  
By repeating this exercise for each combination of $\dxr$ and redshift, 
we follow the growth of $\lmfp$ and its dependence on $\dxr$.

\begin{figure*}
\centerline{
\setlength{\epsfxsize}{0.5\textwidth}
\centerline{\epsfbox{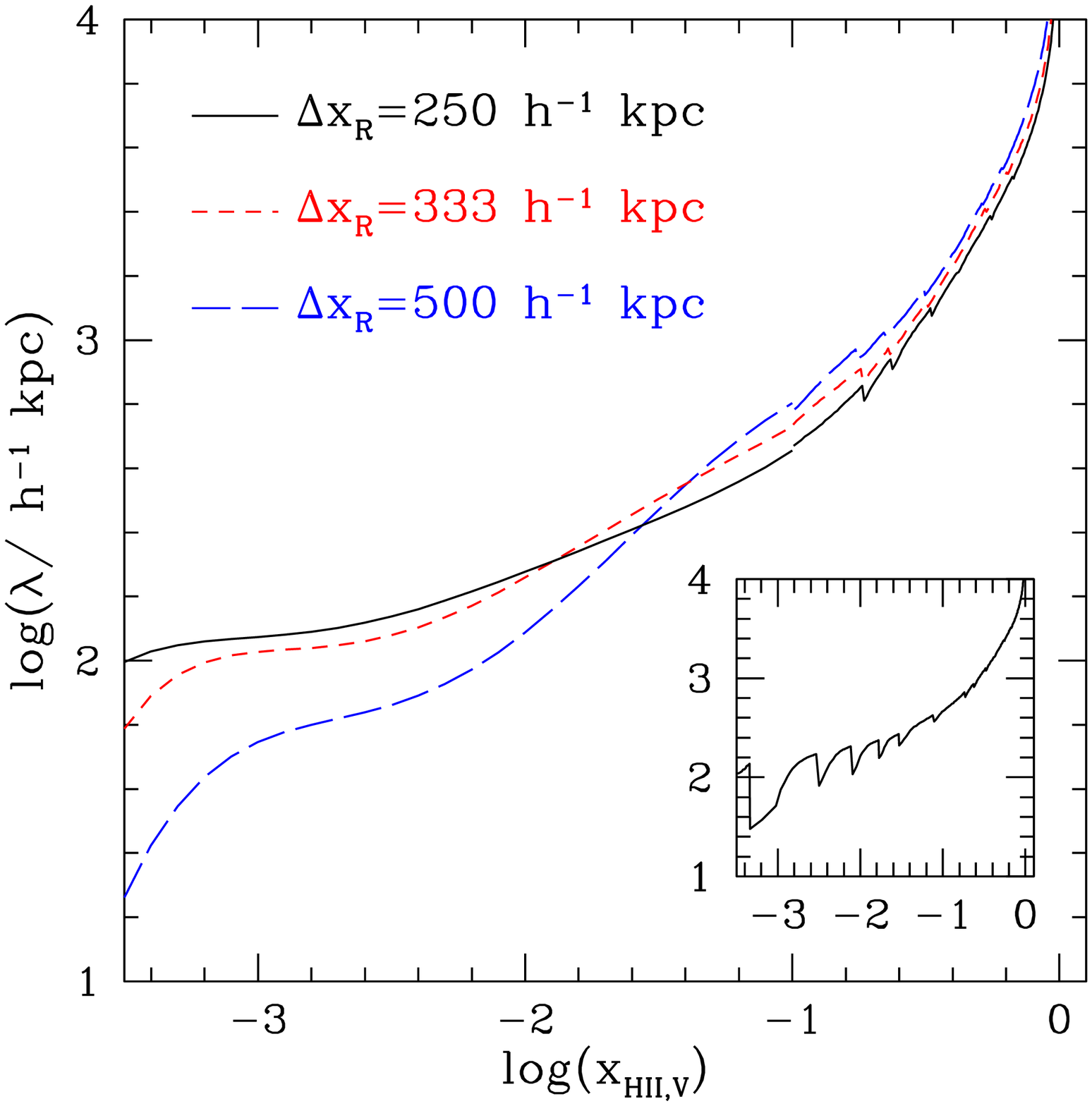}}
}
\caption{The ionizing mean free path $\lmfp$ from sources as a function 
of the volume-weighted ionized hydrogen fraction, computed using three 
different values of $\dxr$.  The larger panel uses 6th-order polynomial 
fits to the simulated trends for $\xhiiv<$10\%, while the inset panel 
shows the full original curve for $\dxr = 250 \hkpc$.  At early times, 
higher resolution leads to higher $\lmfp$ because photons ``tunnel" more 
effectively from sources to voids.  At late times, higher spatial 
resolution leads to lower $\lmfp$ because the small overdensities that 
dominate the IGM opacity are better resolved.}
\label{fig:raycast}
\end{figure*}

We show the resulting trends of mean free path versus ionized volume 
fraction in Figure~\ref{fig:raycast}.  During the early stages of 
reionization, when the neutral fraction is above 99\%, higher 
resolution leads to higher $\lmfp$.  This is the result that we 
expected: At higher spatial resolution, the tendency of photons to bore 
holes through the IGM from sources to voids is better resolved, leading 
to a higher $\lmfp$ at a given ionization state.  In fact, 
Figure~\ref{fig:raycast} indicates that even our highest-resolution 
computation does not fully capture all the relevant substructure, 
suggesting that the characteristic size of the tunnels is smaller than 
250 comoving $\hkpc$.  It is likely that using a spatial resolution that 
is comparable to the virial radius of the dominant haloes is necessary for
full resolution convergence.  Unfortunately, this requirement is roughly 
a factor of 10 higher than what we have achieved here, and would require 
either a much smaller cosmological volume (thus increasing cosmic variance 
and biasing effects;~\citealt{bar04}), an adaptive RT mesh, or a less 
accurate technique.  Nevertheless, the trend indicates that reionization 
proceeds rapidly from overdensities into voids before mopping up the 
filaments.  It also reinforces the need for spatial resolutions that are 
much higher than 1 comoving $\hmpc$ when studying the topology of 
reionization.

As reionization proceeds, the trend of $\lmfp$ versus resolution inverts 
so that higher resolution leads to lower $\lmfp$.  This occurs 
as the topology changes from one in which ionized tunnels thread an 
otherwise opaque IGM to one in which increasingly isolated overdensities 
are separated from each other by regions that are already reionized.  
The remaining neutral regions dominate the IGM opacity and shrink as 
reionization proceeds.  Higher spatial resolution prevents them from 
being averaged with the low recombination rates in neighboring voids.  
At this stage, reionization proceeds from the voids back into regions of 
increasing density and decreasing spatial scale~\citep{mir00}.  Our 
simulations suggest that this topology could dominate by the time the 
neutral fraction drops below 50\%, and possibly earlier.

\section{Comparison to Observational Constraints}
\label{sec:wmap}
The goal of the present work is to take a hydrodynamical model that has
enjoyed considerable success in accounting for observational constraints 
from the post-reionization Universe and to ask how reionization proceeds 
in this model.  Because we have introduced no extra physical 
assumptions save for the choice of ionizing escape fraction and have tuned
this to achieve reionization by $z=6$, it is of interest to compare our 
results with additional observational constraints from the reionization 
epoch, specifically Thomson optical depth measurements from the 
\emph{Wilkinson Microwave Anisotropy Probe} (WMAP), and observations of 
the Lyman-$\alpha$ forest.  In this Section we show that (1) our 
simulated optical depth to Thomson scattering underproduces the observed 
value and (2) our simulations overproduce the mean ionizing intensity at 
$z=6$.  We discuss the implications of these findings.

\subsection{The Integrated Optical Depth to Electron Scattering}
\label{sec:tau_es}

\begin{figure*}
\centerline{
\setlength{\epsfxsize}{0.5\textwidth}
%\centerline{\epsfbox{data/fig.tau_es_ng64.ps}}
\centerline{\epsfbox{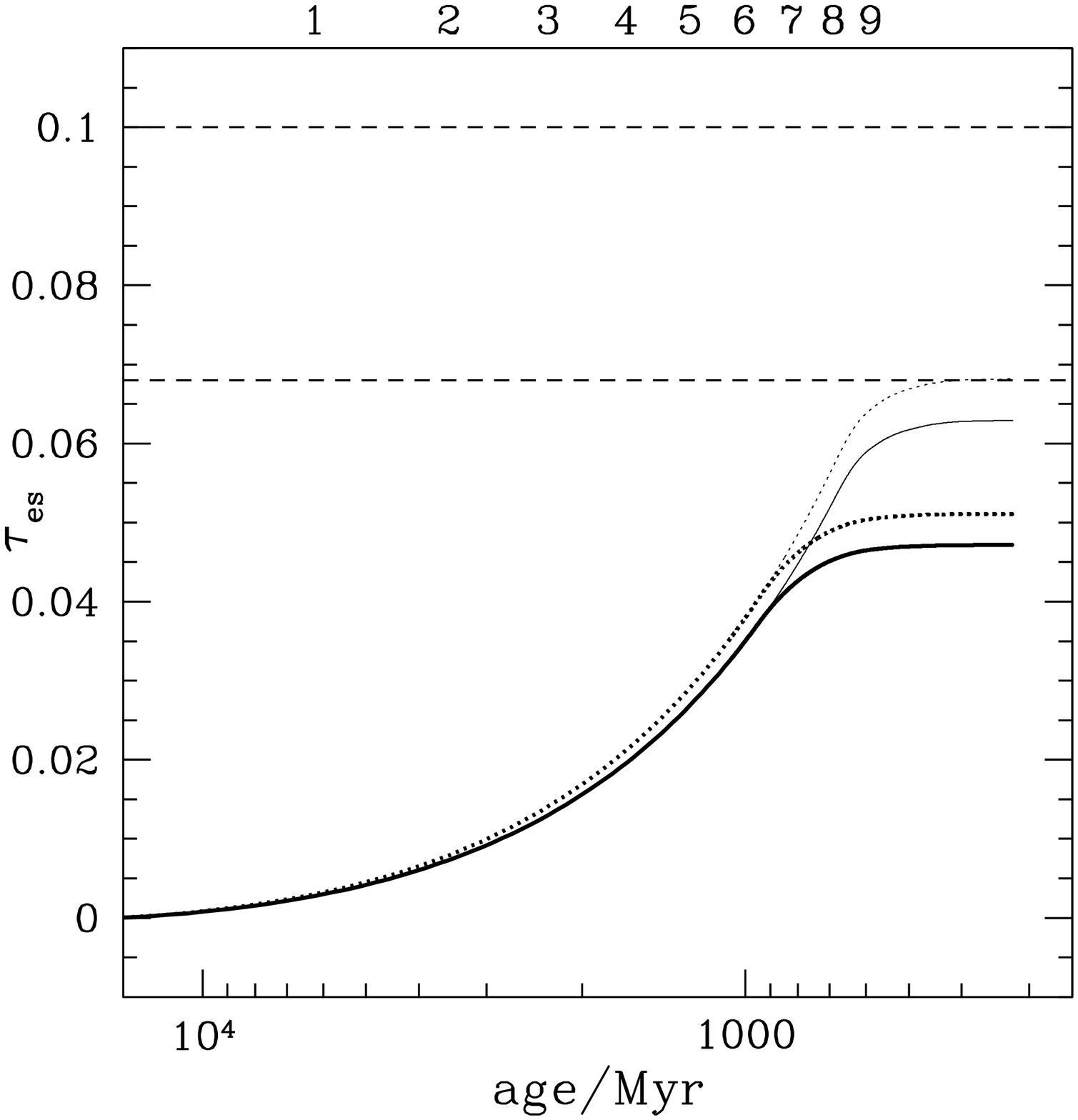}}
}
\caption{The integrated optical depth to Thomson scattering $\tes$ as a 
function of the age of the Universe (bottom axis) and redshift (top axis).
Thick curves correspond to $\dxr=250\hkpc$, $\fesc=0.13$, and fiducial 
accuracy while thin curves correspond to $\dxr=500\hkpc$, $\fesc=1$, and 
the fast scheme (Section~\ref{sec:radtx_sim}).  Solid curves show the result
from pure hydrogeon reionization, and dotted curves include the contribution 
of Helium assuming that Helium is doubly ionized after $z=3$ and singly 
ionized with $\nheii/\nhe = \nhii/\nh$ for $z>3$.  The dashed lines indicate 
the 68\% confidence intervals for $\tes$ arising from combining WMAP-5 
with distance measurements from Type Ia supernovae and baryon acoustic 
oscillations, $\tes=0.084\pm0.016$~\citep{hin09,kom09}.  Our 
simulations do generate enough ionizing photons to match the observed 
constraint on $\tes$, but our fiducial choice of $\fesc$ may underproduce 
it.}
\label{fig:tau_es}
\end{figure*}

In Figure~\ref{fig:tau_es}, we show the variation of the integrated 
optical depth to Thomson scattering $\tes$ with redshift.  The thick 
solid curve corresponds to pure hydrogen reionization at our highest
spatial resolution and accuracy, and indicates an integrated optical
depth of 0.047, 0.023 below the 1-$\sigma$ observational constraints.
Accounting for Helium reionization (thick dotted curve; see caption
for details) boosts $\tes$ to 0.051.  The discrepancy between the
observed and simulated $\tes$, which reflects the history of the
volume-averaged electron number density, suggests that reionization
occurs too suddenly or too late in our calculations.

A number of effects could explain this discrepancy.  On the numerical
side, our hydrodynamic simulation may not resolve all of the relevant
star formation at early times (Figure~\ref{fig:ion_mhalo}).  Increased
star formation at early times could start reionization earlier and 
boost $\tes$ without changing $\xhiv$ at late times, when the ionizing
background is dominated by more massive
haloes~\citep[Figure~\ref{fig:ion_mhalo};][]{ili07}.  The role of 
early star formation could be further enhanced through a more detailed
treatment of the ionizing escape fraction, $\fesc$.  Currently, we have 
set $\fesc$ to a time-invariant value of 13\% in order to obtain the 
end of reionization at $z\approx6$.  However, this is a free parameter 
and could be larger at earlier times~\citep[e.g.,][]{wis09}.  
Raising $\fesc$ to 1 (thin curves) brings our simulated optical depth 
into agreement with the observed 68\% confidence interval, but makes 
the redshift of overlap $z=7.90$.  Note that this change slightly 
modifies the qualitative topology of reionization, but does not make 
it purely IO (Figure~\ref{fig:zreion_sys}).  Third, the finite spatial 
resolution affects $\tes$ because reionization occurs more rapidly at 
higher $\dxr$ (see Section~\ref{sec:topo}).  For radiative transfer 
grid resolutions of $\Delta x = 500, 333, 250 \hkpc$, $\xhiv$ drops 
below $10^{-3}$ at $z=$ 5.97, 6.05, and 6.20, respectively.  Hence 
the choice of $\dxr$ introduces an uncertainty of approximately 
$\Delta z=$0.2.  Finally, our small cosmological volume delays 
reionization by $\Delta z=0.1$ because of the lack of long-wavelength 
density fluctuations~\citep{bar04} and introduces a random uncertainty 
of $\Delta z=1$ owing to cosmic variance~\citep{bar04,ili06a}, which
can be translated to an effect on $\tes$ via $d \tes / d \zreion
\approx 0.008$.  In short, the systematic uncertainties in our
reionization history owing to numerical effects and assumptions are
sufficient to account for the missing optical depth.

But there are further observational uncertainties related to the
observed $\tes$.~~\citet{shu08} have argued that the systematic
uncertainties in computing the residual electron fraction leftover
after recombination lead to systematic uncertainties in $\tes$ of 
order $\sim0.01$.  If true, then this could reduce the amount of 
Thomson scattering that galaxies are responsible for producing, 
bringing our simulations into better agreement with observations.
The unknown time dependence of reionization also renders $\tes$ 
uncertain (although~\citealt{mor08} have argued that realistic 
reionization histories systematically increase $\tes$ over the 
instantaneous reionization value, which would further increase 
the discrepancy with our simulations).

Finally, it is also possible that reproducing $\tes$ in large 
cosmological volumes requires additional physical processes such as 
self-regulation of star formation in low mass haloes~\citep{ili07},
Population III stars~\citep{tra07,shi08}, X-rays from early black
holes~\citep{shu08} or supernovae~\citep{oh01}, or primordial 
magnetic fields~\citep{sch08}.  In future work, we will use radiative 
hydrodynamic simulations of reionization to study self-regulation; 
however, for the present, our goal is to study how reionization 
follows from our existing baryonic physics treatments with no 
additional assumptions.

\subsection{The Hydrogen Ionization Rate at $z=6$}
\label{sec:gamma}

The solid blue curve in the bottom panel of Figure~\ref{fig:xHI_vs_t} shows 
how the volume-averaged hydrogen ionization rate per hydrogen atom
$\Gamma_{-12}\equiv\Gamma_{\mathrm{H\,I}}/10^{-12}$ varies with redshift in 
our calculation.  At $z=6$, we find $\log(\Gamma_{-12}) = 1$, whereas 
observations of the Lyman-$\alpha$ forest indicate 
$\log(\Gamma_{-12})< -0.6$~\citep[triangle with arrow;][]{bol07,srb08}.
Relatedly, our fiducial simulation yields a volume-averaged neutral fraction 
at $z=6$ of $2.4\times10^{-6}$, 50--100 times smaller than the observed 
2--3$\times10^{-4}$~\citep{fan06}.  

These large discrepancies could result either 
from $\fesc$ being too high or from the opacity in the reionized IGM being 
too low.  The fact that reionization occurs more rapidly at higher spatial 
resolution (Figure~\ref{fig:zreion}) indicates that we are forced to choose 
$\fesc$ artificially large in order to achieve overlap by $z=6$.  
Meanwhile, the fact that the IGM opacity at a given ionization state 
increases with decreasing $\dxr$ (Figure~\ref{fig:raycast}) indicates that 
our simulated IGM opacity at $z=6$ is artificially low.  Both of these 
effects indicate that increasing our spatial resolution or incorporating a 
subgrid treatment for photoionization of structures below our spatial
resolution~\citep[e.g.,][]{cia06,tra07} would improve the agreement with 
the observed $\Gamma_{-12}$.

We may estimate how much our $\Gamma_{-12}$ would improve with resolution
by considering the likely impact of structures below our spatial resolution.  
Under ionization equilibrium, the emissivity and ionization rate are related 
by
\begin{equation}
\label{eqn:ioneq}
\eta = \chi \Gamma / 4 \pi \sigma.
\end{equation}
Here, $\eta=2.694\times10^{-24} (\fesc/0.13)$ is our simulated volume-averaged 
comoving emissivity at $z=6$ in s$^{-1}$ cm$^{-3}$ Sr$^{-1}$; $\chi$ is the 
opacity to ionizing photons in cm$^{-1}$; and $\sigma$ is the neutral hydrogen 
cross section at the Lyman limit in cm$^{-2}$.  At $z=6$, the IGM opacity is 
dominated by Lyman limit systems, hence we may approximate $\chi$ as the 
reciprocal of the mean free path to Lyman limit systems.  Extrapolating the 
number density found by~\citet{sto94}, this is 84 comoving Mpc in our 
cosmology at $z=6$.  \citet{fau08} have estimated the additional impact of 
structures at lower column densities, finding that the mean free path is 
roughly $85[(1+z)/4]^{-4}$ proper Mpc (or 63 comoving Mpc at $z=6$).  
Folding these estimates into Equation~\ref{eqn:ioneq}, we find that 
resolving Lyman limit systems could reduce $\Gamma_{-12}$ from $\approx10$ 
to 2.  This is still larger than the observed upper limit of 
0.3~\citep{bol07,srb08}, suggesting that our escape fraction may still 
be too large.  However, given that increasing the spatial resolution 
also accelerates reionization (Figure~\ref{fig:zreion}), this observation 
reinforces our previous conclusion that decreasing both $\dxr$ and 
$\fesc$ without changing our star formation prescription would yield 
improved agreement with the observations.

Of course, while it is possible that these discrepancies owe to a 
combination of numerical effects and systematic uncertainties in the 
observed optical depth to Thomson scattering, a more interesting 
possibility is that they indicate a failing of the input physics.  
For example, detailed comparisons with observations of galaxies and 
the IGM in the post-reionization Universe have prompted us to assume 
that low-mass galaxies generate strong outflows (Section~\ref{sec:intro}).  
These outflows dramatically suppress the ionizing emissivity at early 
times, leading to a relatively late reionization epoch and a low $\tes$.  
As a reference point, in our simulations, 90\% of the ionizing photons 
that are emitted by $z=6$ are emitted after $z=9$.  Weaker outflows at 
early times would boost early star formation rates and hence $\tes$.  
A stronger scaling between ionizing luminosity and metallicity over 
what is found in the~\citet{bc03} models would enhance the ionizing 
light to mass ratios, as would including a treatment for an evolving 
initial mass function (i.e., Population III star formation), or the 
inclusion of mini-quasars.  Finally, an ionizing escape fraction that 
decreases with decreasing redshift or increasing halo mass would 
enhance the ionizing luminosity into the IGM without changing the star 
formation history.  However, until the numerical issues are more fully 
investigated, we cannot make robust conclusions about the need for new 
or enhanced sources of ionizing photons at the earliest epochs.

In summary, the failure of our simulations to reproduce simultaneously 
the observed optical depth to Thomson scattering and the hydrogen
ionization rate at $z=6$ could owe to observational uncertainties, 
numerical effects, or inadequate physics treatments.  On the other 
hand, given that our simulations have previously been shown to 
reproduce a wide variety of observations of the post-reionization 
Universe and that the only additional physical assumption that we 
have introduced is the choice of ionizing escape fraction, the level 
of agreement that we find is encouraging.  Our results support the
idea that the processes governing star formation before and after 
reionization may not have been very different~\citep{dav06}.

\section{Discussion}
\label{sec:disc}
% Actual Discussion
The tendency for I-fronts to bypass filaments and escape directly into 
voids occurs generically in calculations that assume static emissivity 
and density fields (as we did in~\citealt{fin09}).  Given high enough 
spatial resolution, such calculations invariably produce $\hii$~regions 
with ``butterfly wing" morphologies~\citep[e.g.,][]{cia01,ili06b,tas06}.  
In addition, this effect has recently been found in radiative hydrodynamic
simulations of the first stages of star formation~\citep{abe07,wis08}.
This was on much smaller scales ($<1 \hmpc$) than the cosmological scales
that we are interested in.  Nonetheless, one may understand the result in 
both cases as a competition between the timescale for photons to bore 
tunnels through the IGM that connect sources with voids on the one hand 
versus the timescale for filaments to self-ionize and the timescale for 
I-fronts to burn directly through filaments on the other.  If photons 
leak into voids more rapidly than the filaments can either be ionized 
or self-ionize, then the voids will reionize before the filaments.

\begin{figure*}
\centerline{
\setlength{\epsfxsize}{0.5\textwidth}
\centerline{\epsfbox{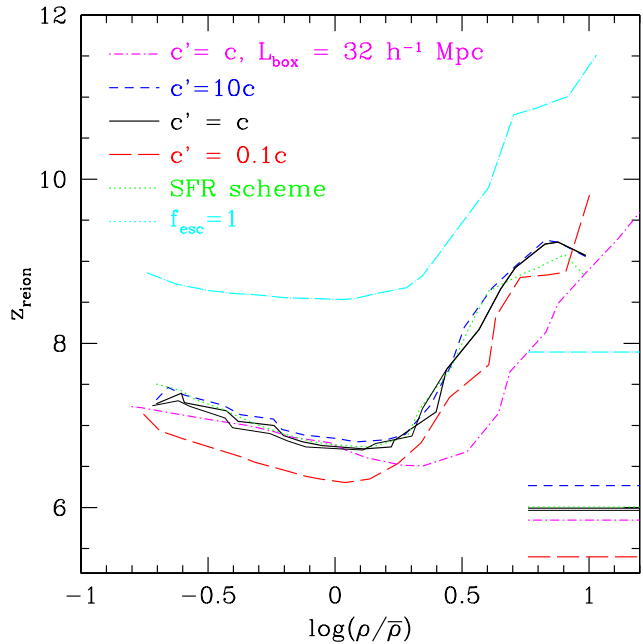}}
}
\caption{Systematic effects on the median reionization redshift as a 
function of overdensity.  The horizontal segments on the right indicate the 
corresponding overlap redshifts.  Short-dashed blue, solid black, and 
long-dashed red curves show the results assuming three different values $c'$ 
for the speed of light.  The dotted green curve shows the results when
the emissivity is derived from the star formation rates.  The dot-dashed
magenta curve uses a cosmological volume eight times as large as our fiducial
volume.  Because these computations used somewhat less accurate Eddington 
tensors in order to save time, we also include the result at our fiducial 
accuracy (light grey curves).  Changing the speed of light or the volume 
does not qualitatively change the topology of reionization, but it does 
impact the overlap redshift.  Changing how we derive the emissivity field 
does not affect our results either.}
\label{fig:zreion_sys}
\end{figure*}

These timescales are in turn governed by three factors: (1) The 
emissivity field's bias with respect to the baryon density field; (2)
the time evolution of the emissivity field; and (3) the spatial 
resolution.  The impact of the emissivity bias on the topology of 
reionization has been explored elsewhere~\citep{mcq07}; for our purposes 
it suffices to note that a highly biased emissivity field promotes the 
late reionization of filaments by preventing them from self-ionizing.  
The time evolution of the ionizing emissivity matters because if the 
emissivity is relatively high at early times, then the late reionization 
of filaments is suppressed even in the presence of a biased emissivity 
field owing to the lower density contrast in the IGM.  As a demonstration 
of this effect, we show the reionization topology resulting from 
setting $\fesc=1$ in Figure~\ref{fig:zreion_sys} (cyan dot-dashed 
curve).  Comparing this curve to the topology in our fiducial case (solid 
black curve) reveals that the higher reionization redshift is indeed 
associated with a flatter trend of median reionization redshift versus 
overdensity at low densities.  Finally, high spatial resolution also 
promotes the early reionization of voids by increasing the density 
contrast.  The topology presented in this paper results from a particular
combination of emissivity bias and evolution that has not been considered
previously, and it is most clearly visible only at our highest spatial 
resolution (Figure~\ref{fig:xmxv}).

\citet{ili06a} ran post-processed radiative transfer calculations of
cosmological reionization on snapshots obtained from a pure N-body 
simulation.  In contrast to our results, they found a purely IO topology.  
Given that their spatial resolution was comparable to ours, the 
difference between our results must owe to the different emissivity 
fields.  In constructing their emissivity field, they considered only 
haloes more massive than $2.5\times10^9\msun$, which implies an even 
more biased emissivity field than ours (Figure~\ref{fig:ion_mhalo}).  
\citet{cho09} suggested that their purely IO topology might result from 
using high emissivities.  Indeed, the~\citet{ili06a} simulation achieved 
reionization by $z\sim11$, only $\approx260$ Myr after 
the first sources formed at $z=21$, whereas our simulations are tuned to 
produce reionization at $z=6$, roughly 680 Myr after the computations begin 
at $z=14$.  We conclude that their purely IO topology owed to their high
reionization redshift and the resulting lower IGM density contrast.

Subsequent calculations involving more extended reionization 
histories and lower overlap redshifts have also resulted in more nearly 
IO-like topologies~\citep{tra07,ili07}, hence the overlap redshift alone 
is not the only cause of our topology.  In the case of~\citet{tra07}, we
ascribe the difference to our more biased emissivity field.  For example,
comparing our Figure~\ref{fig:ion_mhalo} with their Figure 1, we find that,
at $z=6$, the minimum halo mass above which 60\% of ionizing photons are 
emitted is $10^{9.4}\msun$ for our simulation versus $10^9\msun$ for 
theirs.  Their more extended reionization history also allows for much
of reionization to take place at higher redshift, when density contrasts
are lower.  Our less gradual reionization history owes directly to
our star formation prescription, which suppresses the early star 
formation rate density with respect to models that do not include 
strong outflows (e.g., Figure 3 of~\citealt{opp06}) and leads to a more 
sudden overlap epoch because of the strong dependence of star formation 
rate on halo mass (Figure~\ref{fig:mhalo_sfr}).

The emissivity field of~\citet{ili07} is also less biased than ours, but
this time in two respects.  First, they assume that haloes with masses 
between $10^{8\mbox{--}9}\msun$ are the sites of Population III star 
formation, leading to lower ratios of mass to ionizing luminosity than 
more massive haloes.  Second, they assume that the ionizing luminosity is
proportional to halo mass, which, even in the absence of Population III 
stars, would yield a less-biased emissivity field than ours since our 
ionizing luminosity scales as $M^{1.3}$ (Figure~\ref{fig:mhalo_sfr}).  
With these assumptions, it is not surprising that their filaments are 
able to self-ionize much more efficiently than ours.

The simulations of~\citet{ili07,ili06a} assume that the speed of light 
is infinite whereas ours do not; could this assumption impact the 
topology of reionization? As this question has not been investigated in 
a fully numerical context previously, we have re-computed 
cosmological reionization using $\dxr=500\hkpc$ assuming three different 
values $c'$ for the speed of light: $c/10$, $c$, and $10c$.  To save 
time, we perform these integrations using our less-accurate ``fast" 
scheme for updating the Eddington tensors (Section~\ref{sec:radtx_sim}).  
Figure~\ref{fig:zreion_sys} shows how the trend of median redshift of 
reionization versus overdensity (curves) and the overlap redshift 
(horizontal lines) vary with the speed of light.  The light solid 
curve reflects our fiducial accuracy and is copied from 
Figure~\ref{fig:zreion}.  It essentially overlaps the dark solid curve,
indicating that the impact of the somewhat less accurate Eddington 
tensors in our fast scheme is small compared to the systematic offset 
from using an inaccurate speed of light.  The median redshift of 
reionization is roughly $\Delta z=0.1$ (50 Myr) earlier at all 
densities for $c'=10c$ and 0.2--0.4 (150 Myr) later for $c'=c/10$, 
although in both cases the effects are stronger in the underdense than 
in the overdense regions.  Overlap (defined as the redshift when 
$\xhiv$ drops below $10^{-3}$) occurs $\Delta z=(0.3,0.6)$ or (50,150) 
Myr (early, late) for $c'=(10c,c/10)$.  Evidently, the consequences of 
decreasing the speed of light~\citep[as done by e.g.][]{aub08} are 
more dramatic than the consequences of boosting it, although neither 
approximation changes the qualitative topology of reionization.

\citet{mcq07} investigated a variety of emissivity fields, including
cases that were much more biased than ours (their models S3 and S4).
Although they did not specifically address the onset of OI 
reionization, we may speculate on whether their results would have 
agreed with ours.  Their fiducial model S1 corresponds to a less 
biased field than ours partly because it includes the contribution of 
all haloes down to $10^8\msun$, and partly because they assume a 
constant mass-to-light ratio.  Their model S3 includes a stronger 
scaling between halo mass and luminosity than we find, but it still 
includes the contribution of the low-mass ($<10^{8.5}\msun$) haloes.  
Their model S4 assumes that haloes with masses below 
$4\times10^{10}\msun$ do not form stars, hence it corresponds to an 
extremely biased emissivity field.  However, because they tune the 
emissivities to match the volume-averaged emissivity in their S1 
model, reionization is already well underway before the density 
contrasts are large enough for the filaments to self-shield 
efficiently.  Hence, of their models, the S3 would have been the most 
similar to ours.  Unfortunately, their spatial resolution is slightly 
lower than ours ($367\hkpc$ versus $250\hkpc$), which artificially 
suppresses the density contrast and hence the onset of the OI phase 
(Figure~\ref{fig:xmxv}).  

It is possible that increasing our mass resolution would boost the 
star formation rates in low mass haloes
at early times, leading to a less biased emissivity and more extended
reionization history.  However, even in this case our results would 
remain sensitive to the unknown ionizing escape fraction, which may 
decrease with decreasing mass~\citep{gne08,wis09} or increasing gas 
fraction~\citep{oey07}.  Our finding that the topology of reionization 
depends sensitively on assumptions regarding baryonic physics in 
low-mass haloes emphasizes the need for improved observational 
constraints on the faint end of the luminosity function as well as on 
the escape fraction of ionizing photons.  It also emphasizes the need 
for high-resolution radiative hydrodynamic simulations of galaxy 
formation during the reionization epoch~\citep[e.g.][]{wis09} in 
order to tune the assumptions that go into larger-scale simulations.

% Sources of Uncertainty

Our limited cosmological volume may introduce some uncertainty into
our results.  It has been shown that the lack of long-wavelength 
density fluctuations sampled by our $16\hmpc$ volume delays overlap by 
$\Delta z = 0.1$ and introduces an uncertainty of $\Delta z = 1$ owing to 
cosmic variance~\citep[e.g.,][]{bar04,ili06a}.  While it has not been shown
that these effects qualitatively change the topology of reionization, we
have explored this possibility by computing reionization using snapshots
extracted from a $32\hmpc$ cosmological hydrodynamic simulation.  The
baryonic physics treatments in this simulation are the same as in our 
fiducial volume, as is the assumed ionizing escape fraction.  Its spatial 
resolution is $8\times$ lower, with the result that star formation is 
artificially suppressed in haloes below the 64-particle limit of 
$1.5\times10^9\msun$.  We use a grid of $64^3$ cells, making $\dxr$
identical to the tests in Figure~\ref{fig:zreion_sys}.  The dot-dashed 
magenta curve in Figure~\ref{fig:zreion_sys} shows the resulting 
reionization topology.  Comparing it to the solid black curve, we find 
that the topology is qualitatively unchanged.  In detail, overdense 
regions reionize later owing to the delay of star formation at lower 
mass resolution~\citep[e.g.,][]{spr03b}.  This in turn delays the overlap 
epoch by roughly $\Delta z = 0.2$.  Ideally, we would prefer run this 
test on outputs from an even larger hydrodynamic simulation that used 
the same mass resolution.  Unfortunately, this is not possible with our 
current computing resources.  Nevertheless, the qualitative agreement 
between our $16\hmpc$ and $32\hmpc$ volumes suggests both that our 
topology does not owe to cosmic variance, and that using an even larger 
volume would not change our results.  It would be interesting to 
compute the reionization of a $16\hmpc$ or $32\hmpc$ volume using a 
testbed that is known to yield a purely inside-out reionization 
topology for larger volumes; however, this is beyond the scope of our 
present work.

An additional source of uncertainty in our results is the unknown impact 
of structures on size scales below our spatial resolution such as 
virialized haloes with masses in the range $10^{5\mbox{--}7}\msun$ (i.e., 
minihaloes).  Given that the timescale for ionizing photons to bore holes 
through the IGM into the voids is comparable to the timescale over which 
the Universe emits enough photons to ionize each baryon once, 
minihaloes---which are concentrated in filaments---could tilt the 
competition further in favor of an IOM topology through their ability to 
delay the completion of reionization by up to 
$\Delta z = 2$~\citep[e.g.,][]{bar02,cia06,mcq07}.  Moreover, analytic and 
numerical studies now agree that minihaloes change the topology of 
reionization, decreasing the number of large $\hii$~regions and increasing 
the number of small ones~\citep{mcq07,fur05} at late times.  Although we do 
not actually resolve minihaloes, this finding is completely consistent with 
the trend that we found in Figure~\ref{fig:raycast} whereby the mean free 
path to ionizing photons in the latter stages of reionization is shorter at 
higher spatial resolution owing to the impact of overdense ``holdouts".  
This, however, is in the late stages of reionization ($\xhiiv>0.5$); what 
about during the early stages? Minihaloes cannot substantially affect the 
topology of reionization until the mean free path to absorption by the IGM
is comparable to mean free path to pass within a virial radius of a 
minihalo, which, for a Press-Schechter mass function and our reionization 
history, happens after $z=8$.  By contrast, Figure~\ref{fig:xHI_vs_t} 
indicates that the strength of the ionizing background in voids is already 
comparable to the volume average by $z=9$.  We conclude that minihaloes are 
not likely to prevent the IOM topology because they are not numerous enough 
to obstruct the photons' paths into the voids.

\section{Summary}
\label{sec:conc}
We have used our new moment-based radiative transfer technique to study 
how cosmological reionization proceeds on precomputed grids of density 
and ionizing emissivity spanning a $16^3 h^{-3}$ Mpc$^3$ volume.  We 
extract these grids from a cosmological hydrodynamic simulation that 
resolves all star formation in dark matter haloes more massive than 
$2\times10^8\msun$.  This mass threshold accounts for most of the 
relevant star formation prior to the onset of reionization and all 
of it once reionization is well underway since the nascent ionizing 
background suppresses star formation in haloes less massive than 
$\sim10^9\msun$.

We find that reionization proceeds rapidly from overdense regions
into voids owing to the porous nature of the IGM, with filamentary
structures ionizing last, which we call an inside-outside-middle
(IOM) topology.  This is consistent with what we obtained previously using
less sophisticated methods~\citep{fin09}.  In our models this IOM
topology arises because filaments cannot self-ionize, since haloes
less massive than $\sim$a few$\times10^8\msun$ do not form stars 
efficiently.  While this cutoff may owe to resolution limitations 
in our simulation prior to the onset of reionization, it is an 
expected consequence of Jeans mass suppression in reionized regions 
and the primordial cooling floor at $10^4$~K.

The mean density of ionized regions $\xmxv$ drops from high values at 
early times to the cosmological mean density following the epoch of 
overlap, as expected given that sources lie predominantly in overdense 
regions.  At higher spatial resolution, the tendency for photons to 
``leak" directly into voids suppresses $\xmxv$ more rapidly so that,
at our highest resolution, the mean density of ionized regions drops
below the cosmological mean density once the volume-averaged ionized
hydrogen fraction $\xhiiv$ surpasses 50\%.  This leaking effect also
manifests as a tendency for the mean free path for ionizing photons 
$\lmfp$ to increase at higher spatial resolution for $\xhiiv < 10$\%.  
At later stages, $\lmfp$ decreases at higher resolution because the 
dense condensations that dominate the IGM opacity at late times are
better resolved.

Our reionization history underpredicts the integrated optical depth to
Thomson scattering and overpredicts the strength of the ionizing
background at $z=6$.  The low optical depth could indicate that 
reionization begins too late and that $\fesc$ should be increased, while
the high ionizing background suggests that the IGM opacity at $z=6$
is too low.  Increasing $\fesc$ to unity leads to better agreement
with the observed optical depth while exacerbating the discrepancy
with the observed ionizing background at $z=6$.  An interesting
alternative would be to have $\fesc=1$ for small haloes 
while leaving it at $\leq10$\% for larger haloes.
This would increase the electron fraction at early times, when the
ionizing background is dominated by low-mass haloes, while leaving
it unchanged at late times, when the haloes below $5\times10^8\msun$
no longer form stars.  This would be analogous to models considering
Population III star formation and self-regulation of star formation
in low-mass haloes, which have been shown to yield good agreement 
with WMAP observations~\citep{ili07}.

The qualitative topology of reionization is robust to our choice of
spatial resolution, cosmological volume, and ionizing escape 
fraction even though each of these affects the overlap redshift.  
Increasing the spatial resolution delays reionization in 
overdense regions owing to more effective I-front trapping while 
advancing the reionization of voids owing to more efficient 
``tunneling" of photons through small-scale soft spots in the IGM.  
The latter effect also causes overlap to occur sooner.  Doubling the 
length of our simulation volume from $16$ to $32\hmpc$ delays 
overlap by $\Delta z\approx 0.1$ because the delayed onset of star 
formation at lower mass resolution wins over the tendency for overlap 
to occur earlier in larger periodic volumes.  The topology, however, 
remains essentially unchanged.  Increasing the ionizing escape 
fraction to unity causes overlap to occur sooner by $\Delta z\approx 2$.
Additionally, although the filaments are still the last regions to 
ionize in this scenario, they do so somewhat sooner with respect to 
the voids owing to the decreased density contrast at higher redshift.  
This suggests that enhancing the ionizing emissivity at early times by, 
for example, increasing the ionizing escape fraction from low-mass halos 
would lead to a more IO-like topology.

We have also explored how approximations regarding the speed of light 
and the accuracy of the Eddington tensors impact reionization.  Broadly,
we find that, while these approximations do not qualitatively change the 
topology of reionization, they do affect the overlap redshift.  We find
that (increasing/decreasing) the speed of light by a factor of 10 
(advances/delays) overlap by (50/150) Myr or $\Delta z=$ (0.3/0.6).  At 
a resolution of $\dxr = 250\hkpc$, using Eddington tensors derived in 
the optically thin limit delays reionization by 40 Myr or 
$\Delta z= 0.2$, primarily by delaying the reionization of filaments.  
These results emphasize the need to avoid such physical approximations 
when an accurate calculation of reionization is desired, and also 
highlight an important feature of our accurate RT code that enables us 
to directly check the impact of such approximations.

Simulations of reionization are in their infancy.  Ideally, one
would like a single model to span first stars calculations on sub-kpc
scales within cosmological volumes that encompass the largest
ionization bubbles at overlap.  Our current codes cannot do so, but
future improvements such as an adaptive RT mesh and its incorporation
into the hydrodynamical evolution will bring us closer to this goal.
Meanwhile, we can still gain key insights into the topology of
reionization, and better understand how one must compute reionization
accurately.  This will set the framework for understanding the
rapidly advancing observations of this final cosmic frontier.

\section*{Acknowledgements}
As this work was heavily informed by conversations with the majority of 
the reionization community, it would be difficult to acknowledge everyone 
who contributed.  However, we would like to thank T.\ Abel, M.\ Alvarez,
J.\ Bolton, M.\ Haehnelt, I.\ Iliev, A.\ Lidz, A.\ Maselli, A.\ Pawlik, 
J.\ Schaye, and H.\ Trac in particular for advice and encouragement.  We 
are additionally grateful to T.\ Abel for suggesting our technique for 
switching from optically thick to optically thin Eddington tensors.  
Finally, we thank the anonymous reviewer for comments that improved the 
manuscript.  Our cosmological hydrodynamic simulation was run on the 
Xeon Linux Supercluster at the National Center for Supercomputing 
Applications, and many of our radiative transfer computations were run 
using the University of Arizona's SGI Altix 4700 ``Marin" as well as 
its Xeon cluster ``Ice".   F\"O acknowledges support from NSF grant 
AST 07-08640.  Support for this work was provided by the NASA 
Astrophysics Theory Program through grant NNG06GH98G, as well as 
through grant number HST-AR-10647 from the SPACE TELESCOPE SCIENCE 
INSTITUTE, which is operated by AURA, Inc. under NASA contract 
NAS5-26555.  Support for this work, part of the Spitzer Space Telescope 
Theoretical Research Program, was also provided by NASA through a 
contract issued by the Jet Propulsion Laboratory, California Institute 
of Technology under a contract with NASA.


\begin{thebibliography}{99}
\frenchspacing
\bibitem[Abel et al.(2007)]{abe07} Abel, T., Wise, J.~H., 
\& Bryan, G.~L.\ 2007, \apjl, 659, L87 
\bibitem[Ahn et al.(2009)]{ahn09} Ahn, K., Shapiro, P.~R., 
Iliev, I.~T., Mellema, G., \& Pen, U.-L.\ 2009, \apj, 695, 1430 
\bibitem[Aubert \& Teyssier(2008)]{aub08} Aubert, D., \& 
Teyssier, R.\ 2008, \mnras, 387, 295 
\bibitem[Barkana \& Loeb(2002)]{bar02} Barkana, R., \& Loeb, A.\ 2002, \apj, 578, 1 
\bibitem[Barkana \& Loeb(2004)]{bar04} Barkana, R., \& 
Loeb, A.\ 2004, \apj, 609, 474 
\bibitem[Bolton \& Haehnelt(2007)]{bol07} Bolton, J.~S., \& 
Haehnelt, M.~G.\ 2007, \mnras, 382, 325 
\bibitem[Bouwens et al.(2007)]{bou07} Bouwens, R.~J., 
Illingworth, G.~D., Franx, M., \& Ford, H.\ 2007, \apj, 670, 928 
\bibitem[Bruzual \& Charlot(2003)]{bc03} Bruzual, G. \& Charlot, S. 2003,
MNRAS, 344, 1000
\bibitem[Chabrier(2003)]{cha03} Chabrier, G.\ 2003, PASP, 115, 763
\bibitem[Choudhury et al.(2009)]{cho09} Choudhury, T.~R., 
Haehnelt, M.~G., \& Regan, J.\ 2009, \mnras, 394, 960 
\bibitem[Ciardi et al.(2001)]{cia01} Ciardi, B., Ferrara, A.,
Marri, S., \& Raimondo, G.\ 2001, \mnras, 324, 381
\bibitem[Ciardi et al.(2006)]{cia06} Ciardi, B., Scannapieco, E., Stoehr, 
F., Ferrara, A., Iliev, I.~T., \& Shapiro, P.~R.\ 2006, \mnras, 366, 689 
\bibitem[Dav{\'e} et al.(2006)]{dav06} Dav{\'e}, R., 
Finlator, K., \& Oppenheimer, B.~D.\ 2006, \mnras, 370, 273 
\bibitem[Dav{\'e} et al.(2008)]{dav08} Dav{\'e}, R., 
Oppenheimer, B.~D., \& Sivanandam, S.\ 2008, \mnras, 391, 110 
\bibitem[Fan et al.(2006)]{fan06} Fan, X., et al.\ 2006, \aj, 132, 117 
\bibitem[Faucher-Gigu{\`e}re et al.(2008)]{fau08} 
Faucher-Gigu{\`e}re, C.-A., Lidz, A., Hernquist, L., 
\& Zaldarriaga, M.\ 2008, \apj, 688, 85 
\bibitem[Finlator \& Dav{\'e}(2008)]{fin08} Finlator, K., \& 
Dav{\'e}, R.\ 2008, \mnras, 385, 2181 
\bibitem[Finlator et al.(2009)]{fin09} Finlator, K., 
{\"O}zel, F., \& Dav{\'e}, R.\ 2009, \mnras, 393, 1090 
\bibitem[Finlator et al.(2007)]{fin07} Finlator, K., 
Dav{\'e}, R., \& Oppenheimer, B.~D.\ 2007, \mnras, 376, 1861 
\bibitem[Furlanetto \& Oh(2005)]{fur05} Furlanetto, S.~R., \& 
Oh, S.~P.\ 2005, \mnras, 363, 1031 
\bibitem[Gnedin \& Ostriker(1997)]{gne97} Gnedin, N.~Y., \& Ostriker, J.~P.\
1997, \apj, 486, 581
\bibitem[Gnedin(2000)]{gne00} Gnedin, N.~Y.\ 2000, \apj, 535, 530 
\bibitem[Gnedin et al.(2008)]{gne08} Gnedin, N.~Y., Kravtsov, 
A.~V., \& Chen, H.-W.\ 2008, \apj, 672, 765 
\bibitem[Haiman et al.(1997)]{hai97} Haiman, Z., Rees, M.~J., 
\& Loeb, A.\ 1997, \apj, 476, 458 
\bibitem[Haardt \& Madau(2001)]{haa01} Haardt, F. \& Madau, P. 2001, in proc.
XXXVIth Rencontres de Moriond, eds. D.M. Neumann \& J.T.T. Van.
\bibitem[Hinshaw et al.(2009)]{hin09} Hinshaw, G., et al.\ 
2009, \apjs, 180, 225 
\bibitem[Iliev et al.(2006a)]{ili06a} Iliev, I.~T., Mellema, 
G., Pen, U.-L., Merz, H., Shapiro, P.~R., 
\& Alvarez, M.~A.\ 2006, \mnras, 369, 1625 
\bibitem[Iliev et al.(2006b)]{ili06b} Iliev, I.~T., et al.\ 2006, 
\mnras, 371, 1057 
\bibitem[Iliev et al.(2007)]{ili07} Iliev, I.~T., Mellema, 
G., Shapiro, P.~R., \& Pen, U.-L.\ 2007, \mnras, 376, 534 
\bibitem[Kennicutt(1998a)]{ken98a} Kennicutt, R.~C. 1998, ApJ, 498, 541
\bibitem[Kennicutt(1998a)]{ken98b} Kennicutt, R.~C., Jr.\ 1998,
\araa, 36, 189 
\bibitem[Komatsu et al.(2009)]{kom09} Komatsu, E., et al.\ 
2009, \apjs, 180, 330 
\bibitem[Lee et al.(2008)]{lee08} Lee, K.-G., Cen, R., Gott, 
J.~R.~I., \& Trac, H.\ 2008, \apj, 675, 8 
\bibitem[Lidz et al.(2009)]{lid09} Lidz, A., Zahn, O., 
Furlanetto, S.~R., McQuinn, M., Hernquist, L., \& Zaldarriaga, M.\ 2009, \apj, 690, 252 
\bibitem[McKee \& Ostriker(1977)]{mck77} McKee, C.~F. \& Ostriker, J.~P.\ 1977,
ApJ, 218, 148
\bibitem[McQuinn et al.(2007)]{mcq07} McQuinn, M., Lidz, A., Zahn, O., 
Dutta, S., Hernquist, L., \& Zaldarriaga, M.\ 2007, \mnras, 377, 1043
\bibitem[Miralda-Escud{\'e} et al.(2000)]{mir00} 
Miralda-Escud{\'e}, J., Haehnelt, M., \& Rees, M.~J.\ 2000, \apj, 530, 1 
\bibitem[Mortonson \& Hu(2008)]{mor08} Mortonson, M.~J., \& Hu, W.\ 2008, \apjl, 686, L53 
\bibitem[Nakamoto et al.(2001)]{nak01} Nakamoto, T., Umemura, 
M., \& Susa, H.\ 2001, \mnras, 321, 593 
\bibitem[Oey et al.(2007)]{oey07} Oey, M.~S., et al.\ 2007, 
\apj, 661, 801 
\bibitem[Oh(2001)]{oh01} Oh, S.~P.\ 2001, \apj, 553, 499 
\bibitem[Okamoto et al.(2008)]{oka08} Okamoto, T., Gao, L., 
\& Theuns, T.\ 2008, \mnras, 390, 920 
\bibitem[Oppenheimer \& Dav{\'e}(2006)]{opp06} Oppenheimer, B.~D., \&
Dav{\'e}, R.\ 2006, \mnras, 373, 1265
\bibitem[Oppenheimer \& Dav{\'e}(2008)]{opp08} Oppenheimer, B.~D., \& 
Dav{\'e}, R.\ 2008, \mnras, 387, 577 
\bibitem[Oppenheimer et al.(2009)]{opp09} Oppenheimer, B.~D., 
Dav{\'e}, R., \& Finlator, K.\ 2009, \mnras, 396, 729 
\bibitem[Schleicher et al.(2008)]{sch08} Schleicher, 
D.~R.~G., Banerjee, R., \& Klessen, R.~S.\ 2008, \prd, 78, 083005 
\bibitem[Shin et al.(2008)]{shi08} Shin, M.-S., Trac, H., 
\& Cen, R.\ 2008, \apj, 681, 756 
\bibitem[Shull \& Venkatesan(2008)]{shu08} Shull, J.~M., \& Venkatesan, A.\
2008, \apj, 685, 1 
\bibitem[Sutherland \& Dopita(1993)]{sut93} Sutherland, R.~S. \& Dopita, M. A.
1993, ApJS, 88, 253
\bibitem[Springel \& Hernquist(2002)]{spr02} Springel, V. \& Hernquist, L.
2002, MNRAS, 333, 649
\bibitem[Springel \& Hernquist(2003a)]{spr03a} Springel, V. \& Hernquist, L.
2003a, MNRAS, 339, 289
\bibitem[Springel \& Hernquist(2003b)]{spr03b} Springel, V., \& Hernquist, L.\ 2003,
\mnras, 339, 312 
\bibitem[Srbinovsky \& Wyithe(2008)]{srb08} Srbinovsky, J., \& Wyithe, S.\ 2008,
arXiv:0807.4782 
\bibitem[Storrie-Lombardi et al.(1994)]{sto94} 
Storrie-Lombardi, L.~J., McMahon, R.~G., Irwin, M.~J., 
\& Hazard, C.\ 1994, \apjl, 427, L13 
\bibitem[Tasitsiomi(2006)]{tas06} Tasitsiomi, A.\ 2006, \apj, 
645, 792 
\bibitem[Thoul \& Weinberg(1996)]{tho96} Thoul, A.~A., \& Weinberg, D.~H.\ 1996,
\apj, 465, 608 
\bibitem[Trac \& Cen(2007)]{tra07} Trac, H., \& Cen, R.\ 2007, \apj, 671, 1 
\bibitem[Trac et al.(2008)]{tra08} Trac, H., Cen, R., 
\& Loeb, A.\ 2008, \apjl, 689, L81 
\bibitem[Wise \& Cen(2009)]{wis09} Wise, J.~H., \& Cen, R.\ 2009, \apj, 693, 984 
\bibitem[Wise \& Abel(2008)]{wis08} Wise, J.~H., \& Abel, T.\ 2008, \apj, 684,
1 
\end{thebibliography}
\end{document}